\documentclass[12pt]{article}
\pdfoutput=1
\usepackage{jheppub}
\usepackage{subcaption}
\usepackage{xspace}
\usepackage{booktabs}
\usepackage{ulem}
\usepackage{listings}
\usepackage{braket}
\usepackage{url}

\title{Event generation for beam dump experiments}

\author[a,b]{Luca Buonocore,}
\author[c]{Claudia Frugiuele,}
\author[d,e]{Fabio Maltoni,}
\author[d]{Olivier Mattelaer,}
\author[b]{Francesco Tramontano}

\affiliation[a]{Physik Institut, Universit{\"a}t Z{\"u}rich, Winterthurerstr.190, 8057 Z\"urich, Switzerland}
\affiliation[b]{Dipartimento di Fisica, Universit\`a di Napoli Federico II and INFN,
  sezione di Napoli, via Cintia, 80126 Napoli, Italy}
\affiliation[c]{Department of Particle Physics and Astrophysics, Weizmann Institute of Science,
Rehovot 7610001, Israel}
\affiliation[d]{Centre for Cosmology, Particle Physics and Phenomenology (CP3),
Universit\'e catholique de Louvain, B-1348 Louvain-la-Neuve, Belgium}
\affiliation[e]{Dipartimento di Fisica e Astronomia, Universit\`a di Bologna and INFN, Sezione di Bologna, via Irnerio 46, 40126 Bologna, Italy}

\preprint{
  \small
  \begin{flushright}
    XX--TH yy/18 \\ CP3--18--70 \\ ZU-TH 46/18
  \end{flushright}
}

\abstract{
A wealth of new physics models which are motivated by questions such as the nature of dark matter, the origin of the neutrino masses and the baryon asymmetry in the universe, predict the existence of hidden sectors featuring new particles. Among the possibilities are heavy neutral leptons, vectors and scalars, that feebly interact  with the Standard Model (SM) sector and are typically light and long lived.  
Such new states could be produced in high-intensity facilities, the so-called beam dump experiments, either directly in the hard interaction or as a decay product of heavier mesons. They could then decay back to the SM or to hidden sector particles, giving rise to peculiar decay or interaction signatures  in a far-placed detector. Simulating such kind of events presents a challenge, as not only short-distance new physics (hard production, hadron decays, and interaction with the detector) and usual SM phenomena  need to be described but also the travel  has to be accounted for as determined by the geometry of the detector. In this work, we describe a new plugin to the {\sc MadGraph5\_aMC@NLO} platform, which allows the complete simulation of new physics processes relevant for beam dump experiments, including the various mechanisms for the production of hidden particles, namely their decays or scattering off SM particles, as well as their far detection, keeping into account spatial correlations and the geometry of the experiment. 
}

\newcommand{\mg}{{\sc MG5aMC}}
\begin{document}
 
 \maketitle
\section{Introduction}

Experiments at the LHC have yet not reported  any sign of physics Beyond the Standard Model (BSM). Nevertheless, the problem of reconciling our description of the fundamental interactions and particles with long-standing problems, such as the matter-antimatter asymmetry in the Universe, the evidence for dark matter from many astrophysical and cosmological observations and the origin of the neutrino masses, becomes ever more pressing. Many ideas have been proposed, some of which addressing one problem at the time, others, more ambitious, providing solutions to two or more open questions at the same time. In this context, a recurrent theme is the hypothesis that a hidden sector involving new light particles, might be  coupled to the Standard Model via, for instance, one of the three portals (scalar, fermion and vector) in a feeble way. Such scenarios  can provide not only dark matter candidates, but also other states, such as heavy neutral leptons, vectors, scalars,  which could be long-lived and also possibly decay back to SM particles.  

To prove the existence and measure the properties of such elusive particles is extremely difficult. The situation is in fact similar to neutrino production and detection~\footnote{In this case, it is somewhat instructive to remind that even though we know a lot about neutrinos properties by now,  $\tau$ neutrinos are still quite unknown; with nine charged current $\tau$ neutrino events identified by the DONUT experiment~\cite{Kodama:2007aa} and 10 by the OPERA experiment~\cite{Agafonova:2018auq}
it is by far the least known of the SM particles.}:   as the energy does not pose a hindrance, one is lead to consider high-intensity facilities and design experimental setups that maximise the rates. In short,  one needs very intense beams and then let such beams cross a heavy and instrumented
target to detect their scattering or, if necessary,  to create a decay tunnel as long as possible to observe their decay products. A first example is NA62~\cite{CortinaGil:2017mqf,Drewes:2018gkc} which can run in a beam dump mode and is expected to collect data in this configuration soon. The DUNE~\cite{Adams:2013qkq} experiment will be operational in  2026 to study neutrino oscillations. As a by-product it could also search for hidden sector particles. The SHiP experiment~\cite{Alekhin:2015byh,Anelli:2015pba} has been designed on purpose to search for  such light and feebly interacting particles originated in interactions of 400 GeV/c protons produced by the CERN
SPS~\cite{SHiP:2018yqc}. More recently, other proposals have been put forward to also exploit  proton collisions at the LHC experiments with detectors placed not very far from the collision points, namely, the CODEX-b~\cite{Gligorov:2017nwh}, MATHUSLA~\cite{Curtin:2017izq,Evans:2017lvd} and FASER~\cite{Kling:2018wct,Feng:2017vli} experiments. 

In the present paper we address the issue of how to efficiently simulate the production of a flux of particles belonging to the hidden sector and their subsequent interactions and/or decays. In the following, we will generically call a Beam Dump Facility (BDF) every experiment where a known flux of primary SM
probes strikes on a fixed target and a detector is placed in an optimal position with respect to the target, with the aim of detecting either neutrinos or a new kind of feebly interacting particles  produced off the primary beam interaction. The case of a detector placed close to a collider experiment such as those proposed in~\cite{Gligorov:2017nwh,Curtin:2017izq,Evans:2017lvd,Kling:2018wct,Feng:2017vli} can be equally treated within our framework without any modifications.
In practice, sensitivity studies of such experiments to new physics phenomena, rely on 
the simulation of two distinct processes, one where the new particles are produced
and the other where the new particles (or their decay products) interact with a detector placed at 
some macroscopic distance, from tens of meters to thousands of kilometers.

The production of a feebly interacting particle in a beam dump can proceed through at least
three phenomenologically different phenomena: i) its prompt production in the high energy
scattering of the primary beam particle with a nucleus in the target; ii) as the result of the decay
of SM particles produced in the primary collision or in the cascade process in the target;
iii) through the bremsstrahlung process of primary or secondary particles in the target.
The detection, on the other hand, will proceed either through the decay in flight of new particle back to visible
SM final states or directly through the scattering with the matter in the detector.

The aim of this paper is to provide an implementation that allows the simulation of the complete chain of subprocesses, from the production to the final detection at a BDF in one go.  Our starting point are {\sc FeynRules}~\cite{Christensen:2008py, Degrande:2011ua, Alloul:2013bka} for the implementation of the new physics model lagrangian and {\sc MadGraph5\_aMC@NLO}~\cite{Alwall:2011uj,Alwall:2014hca}, \mg\ for short, for providing  the necessary short-distance physics elements, the  automatic production of particle-level unweighted events and the framework. To achieve maximal flexibility we provide the implementation as a \mg~ plugin, in line with other recently developed applications~\cite{Artoisenet:2010cn,Artoisenet:2012st,Ambrogi:2018jqj}. 
\begin{figure}[t]
  \centering
  \includegraphics[scale=.4,angle=0]{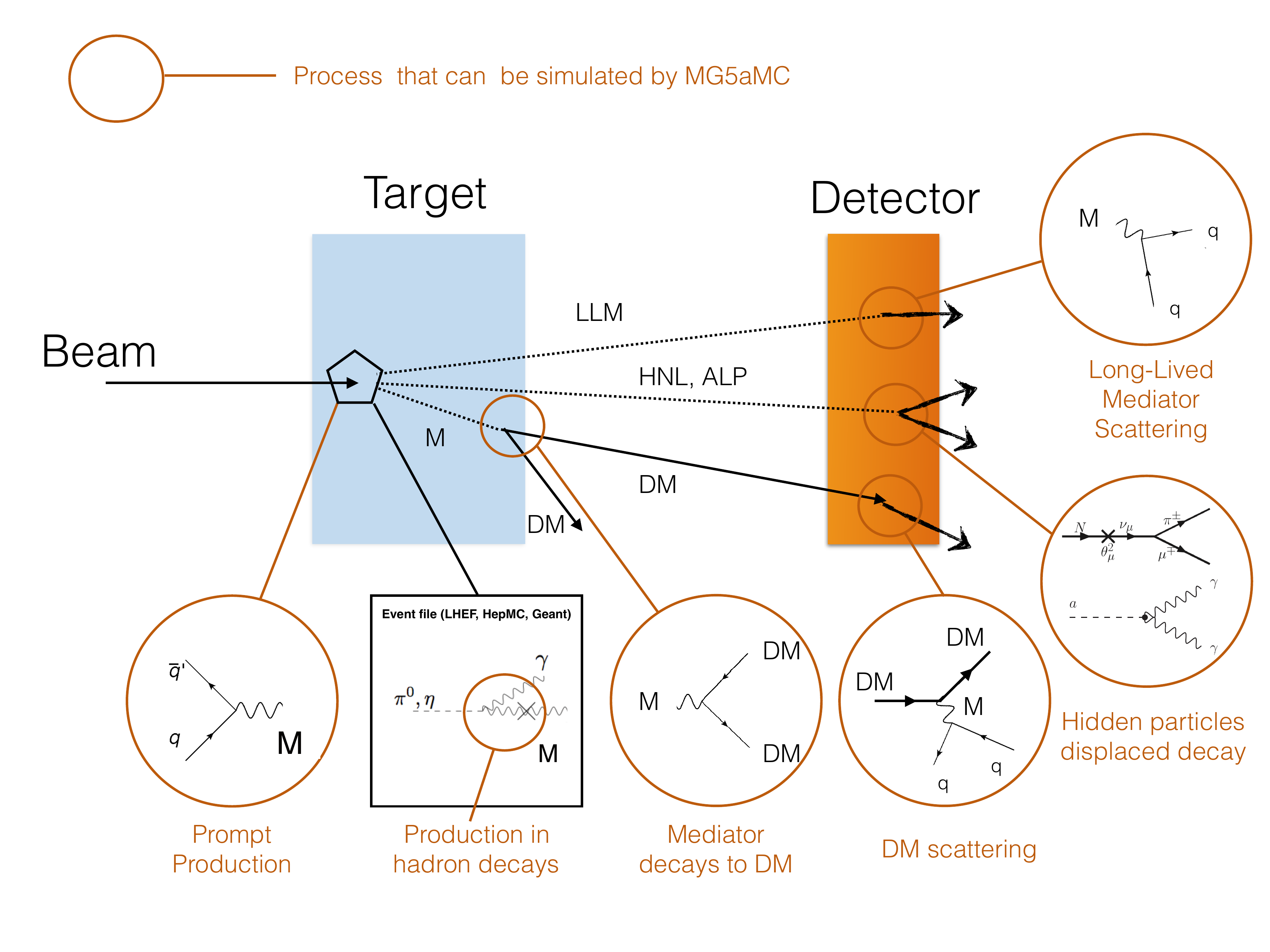}
  \caption{Schematic view of relevant processes that might happen at a BDF
  and can be computed/simulated within \mg~.}\label{fig:MGplot}
\end{figure}
Figure~\ref{fig:MGplot} shows a sketch of the elements of the simulation which are automatically combined in our implementation.  These functionalities are available so that samples of unweighted events in a standard format can be generated in a single step and eventually passed to the simulation of the detector response.
For the rest of the paper we dub the \mg~  plugin for the simulation of hidden particle
effects at beam dump facilities with the short-hand name {\sc MadDump}.

An important aspect of our implementation is that it provides the elements of the simulation that are related to BSM physics in a single framework.  This entails a number of advantages.
First, it eliminates the possibility of making mistakes in the generation or in the combination of event samples for the production and the detection stages.  This is particularly relevant when scanning over the parameters of a BSM model, where, although every step is simple in principle, the combinatorics and the bookkeeping would make the whole construction cumbersome. Second, by using functionalities already present in \mg~it allows to fully automate the scanning over the BSM parameters. Third, once implemented in {\sc FeynRules} and available in the UFO format, the same BSM model and parameter points can be constrained in different contexts within the same framework, for instance, in collider physics using {\sc MadAnalysis5}~\cite{Conte:2012fm,Conte:2014zja,Dumont:2014tja} recasting capabilities or in {\sc MadDM}~\cite{Backovic:2013dpa, Backovic:2015cra, Ambrogi:2018jqj}. 

In {\sc MadDump} the primary flux of Standard Model probes that can generate the hidden particles
has to be provided by the user. As shown in Figure~\ref{fig:MGplot}, it can be either the original beam hitting the target, or the flux of hadrons following the hard interaction interaction that can produce  the hidden particle 
through their decays. In the former case, the user has just to provide
the specific particle code of the probe and its energy in the laboratory frame, while
for the latter case the flux can be given as an event file featuring the decaying particle momenta
and specifying the particle identifier. Note that, for the case of hidden particles generated in the target from meson decays, our approach is more flexible than just directly linking event generators like {\sc Pythia8}~\cite{Sjostrand:2006za,Sjostrand:2014zea} or {\sc HERWIG7}~\cite{Bellm:2015jjp} as it allows, in principle, to later include other effects, such as the cascade production of secondary particles which in some cases could be relevant. {\sc MadDump} is able to handle event files in all formats of the most used event generators.
Another important ingredient is the geometry of the BDF, which is provided by the user in a dedicated file. 

The paper is organised as follows: in section~\ref{Algorithm} we introduce the
algorithms at the core of {\sc MadDump}. In section~\ref{Examples} we present illustrative
examples of possible applications, considering physics cases relevant for the SHiP experiment.
Conclusions and the perspectives of the present work are given
in section~\ref{Conclusions}. In three appendices~\ref{Details1}, \ref{sysunc} and~\ref{Details2} we
provide many details on the numerical techniques employed and the associated uncertainties, while appendix~\ref{Listings}  
documents the scripts that produce the results presented in section~\ref{Examples}.

\section{Approach}
\label{Algorithm}
The first important aspect of our implementation is the idea of considering the beam dump experiment as a two-step process: 
\begin{itemize}
\item {\it Production}: Hidden particle flux generation upon interaction of the beam with the target;
\item {\it Detection}: Interaction of the hidden particles (or their decay products) in the (possibly far-placed) detector.
\end{itemize}
While both steps depend on the details of the new physics model and therefore they have to be considered together, it is possible to factorise the simulation into two independent steps: the results of the {\it Production} phase simulation are used to build a (two-dimensional) parametrisation of the incoming hidden particle flux hitting the detector and leading to different signatures in the detector. By disentangling the {\it Production} from the {\it Detection} phase and the corresponding event generation into two subsequent steps,  the possibility of following the full history, from production to the final signature in the detector, of each event is lost. However, the gain in efficiency in the event generation is enormous, an element which is a key aspect in the simulation of a high-intensity experiment.  

The second important aspect  of {\sc MadDump} is that it has been designed as a plugin of \mg\ . In other words, it heavily relies on already existing modules which are at the core of  \mg~, such as the phase space integration provided by {\sc MadEvent} and the decay package {\sc MadSpin}, integrating them with functionalities that are specifically required for  BDF's, so the various steps of the simulation can be undertaken to obtain the final result in one go.  Among the key new functionalities, we stress 
\begin{itemize}
\item the determination of doubly differential scatter data in the {\it Production} phase of non-standard particle beams  and their support in the {\it Detection} phase;
\item the support of {\sc HepMC} as input format with the aim of making easier the interplay with other Monte Carlo generators like {\sc Pythia8} or {\sc HERWIG7}.
\end{itemize} 

The third aspect is the underlying idea of factorising SM physics from the BSM one, whenever possible. The former, while accessible via standard MC tools, is in general quite involved and needs the modeling of many effects.  Being strongly dependent on the particular experimental setup, a  dedicated simulation of the target and/or detector effects is almost always needed. However, while cumbersome, this part of the simulation does not have any dependence on the new physics model considered and can be taken care once for all.  On the other hand, the new physics short-distance part by definition depends on the details of the model and therefore has to be generated/considered for each different data interpretation. Fortunately, it can be described quite easily from first principles and dealt with by usual or especially developed \mg~ modules. 
\subsection{Production}
In a typical beam dump experiment, a collimated and mono-energetic beam of protons or electrons impinges on a  thick target, at rest in the laboratory frame. A copious number of SM particles is generated both in primary and subsequent secondary interactions inside the target, which is designed to maximise the particles yields. The production of hidden sector particles may proceed according to different mechanisms. 
In the following, we focus on two cases, {\it i.e.}, 
\begin{itemize}
\item prompt production in primary or secondary beam interactions
\item rare meson decays.
\end{itemize}
Without loss of generality, we consider the case of proton beam dump experiment, keeping in mind that other situations can be dealt with by {\sc MadDump} in a completely analogous way.

Depending on the specific BSM physical case (model and parameter point of interest), the prompt production may be described in perturbation theory and it can be treated directly in {\sc MadDump}. In this case, the main input  is the BSM Lagrangian (in the UFO format) which fixes the hidden sector model and its interactions with the SM particles.  For example, in a model where a new massive vector mediator couples to quarks and DM fermionic particles, the main production mechanism resembles Drell-Yan production and decay at fixed target experiments.  This description must, however, be consistent with the typical scales characterising the model. For instance, in the above example the mass of the mediator must be larger then the QCD scale for the computation to be reliable.

As stated above, the main goal of {\sc MadDump} is to handle BSM interactions and then embed them consistently into a complete and modular simulation chain which can take into account the rest of the SM interactions, possibly also using other inputs. In particular, an accurate simulation of the cascade production of hadronic particles, mesons and baryons, is expected to be handled by other MC tools or dedicated simulations, which can fully include parton showers, hadronisation, nuclear effects, meson decays and so on.   This part can be very important if hidden particles are produced in the decays of mesons. In that case, the meson production is assumed to be simulated independently.  {\sc MadDump}, on the other hand, by parsing the event files~\footnote{We remark the importance of  having an easy way to interface {\sc MadDump} with other generators. Indeed, as we argued for the case of meson decay, {\sc MadDump} can take as input the results of other tools in the form of event files. This is the main reason why the {\sc HepMC} format~\cite{Dobbs:2001ck} for the input files was chosen as an option for {\sc MadDump}.} describing the beam-target event takes care of the  decay of mesons into hidden particles, employing for example, an effective field theory approach, which can also be implemented at the level of the UFO.  In this way,  mesons are considered on the same footing of the elementary particles in the model and their decays occur through interaction vertices that can be handled by \mg~ in the usual way.  Examples of both prompt production and meson decay studies are given in section 3. 

Either way, by the hard-interaction or via the decay of mesons, hidden particles are created, which fly out of the
target close to the forward direction. The hidden particles produced during the beam dump, however,  do not form a standard collimated and mono-energetic particle beam. On the contrary, they have a spatial distribution, they are produced in different points inside the volume of the target, and a phase space, i.e., a non trivial four-momenta spread, distribution.  Assuming that the hidden particles travel freely until they eventually enter in the active region of the detector, after macroscopic distances that can go from meters to hundreds or thousands of kilometers, we can describe the beam of hidden particles by means of a multi-differential flux function  
\begin{equation}\label{def_flux}
  \phi(E,\vec{x}) = \frac{dn_\text{DM}}{dEd\vec{x}},
 \end{equation}
where the vector $\vec{x}$ denotes the collection of all the other relevant kinematical variables (angles, spatial distribution of the hidden particles production point within target, etc), but the energy. The flux function  in general not known a-priori and/or in an analytical form since it represents the result of scattering/decay processes in the {\it Production} phase. This distribution is implicitly determined by the simulation of the {\it Production} phase and in turn it can be extracted from a sufficiently large sample of production events. In practice, since the flux depends on the particular BSM model and the specific production mechanism, it cannot be fit it once and for all (as for the proton pdf). An on-the-fly fitting procedure is needed that is fast, robust and flexible.  

\subsection{Detection}

The final detection of the hidden particles might occur according to the two distinct physical processes
\begin{itemize}
\item the hidden particle interacts with the active volume of the detector, resulting in a neutrino-like signature;
\item the hidden particle decays to SM particles inside a dedicated decay tunnel (included of what we dub "detector"), resulting in a "displaced vertex" signature.
\end{itemize}
The interaction of the hidden particle with the detector turns out to be the most difficult part to simulate. One can exploit some approximations and different Monte Carlo techniques to obtain a generator with a satisfactory level of accuracy. On the contrary, as we will discuss later, in the displaced decay case, the same complications do not arise and the situation is much easier to handle. Let us discuss first the interaction case.

\subsubsection{Interactions of hidden particles in the detector}
\label{interaction}
The outcoming flux of hidden particles from the {\it Production} phase,  eq.\eqref{def_flux}, corresponds to the incoming hidden particles flux of the {\it Detection} phase. Our strategy is to parametrise the flux by using {\it Production} event samples and use it as a generalised partonic distribution function (pdf) for the needed computations in the {\it Detection} phase. In doing so, we will also able to parametrise not only the acceptance of the detector but also some of the efficiencies/features that are model dependent.   

The total interaction cross section with the fiducial volume of the detector is obtained by convoluting of the flux function $\phi(E,\vec{x})$ with the elementary cross section $\hat{\sigma}_I$ for the ``partonic'' sub-process
\begin{equation}
\text{HP}+X \rightarrow \text{HP}+X', 
\end{equation}
where $X$ represents the SM matter particle in the detector and "HP" the hidden particle. 
Our implementation is able to handle:
\begin{itemize}
\item elastic electron scattering, $X=e^-$      
\item deep inelastic scattering of nucleons (DIS), $X=u,d,s,c$.
\end{itemize}
The geometrical detector acceptance sets the integration limits in the convolution integral. This is equivalent to introduce a weight function $W(E,\vec{x})$ which is $0$ if the point does not pass the acceptance cut and $1$ otherwise. In this way, it is possible to restore the integration limits to their full ranges. This simple idea is the basis of more sophisticated re-weighting strategies in Monte Carlo integration. We can exploit these techniques with the aim of modeling in a realistic way the detector efficiency. For example, due to its shape and composition, the particles entering the detector may travel a longer or shorter path inside its volume. Correspondingly, the probability that the particles interact inside the detector will be greater or lesser resulting in an efficiency function depending on kinematical variables of the incoming particles. We can effectively describe this effect giving a suitable weight $W(E,\vec{x})$ to each incoming hidden particle particle penalising those which will travel a shorter path. To this aim, we have introduced in our framework the possibility of introducing a re-weighting procedure and we have extensively tested/used it to describe some common detector effects. The user can easily customise the weight function in order to refine the simulation at will. 

Our master formula for the {\it Detection} cross section $\sigma_{D}$ is given by:
\begin{equation}
  \sigma_{D} = \int dE \int d^{n}x \, \phi(E,\vec{x}) \, W(E,\vec{x}) \, \hat{\sigma}_{D}(E).
\end{equation}
The crucial point here is that the partonic cross section $\hat{\sigma}_{D}$ does not depend on the other variables but the energy, as it follows directly from the Lorentz invariance of the interaction among point-like particles. Due to this property we are allowed to formally perform the integral over the $\vec{x}$ vector before performing the convolution with the partonic cross section leading to the introduction of an effective one-dimensional energy pdf:
\begin{equation}
\label{eq:flux_1D}
\tilde{\phi}(E) = \int d\vec{x} \phi(E,\vec{x})W(E,\vec{x}) \implies \sigma_{D} = \int dE \tilde{\phi}(E)
\hat{\sigma}_{D}(E).
\end{equation}
Before proceeding further, we discuss some practical implications of the above formula. The 1D-function $\tilde{\phi}(E)$ can be obtained through a simple 1D fit of the energy histogram of the input DM production events, after having re-weighted them by the weight function $W(E,\vec{x})$. According to eq.\eqref{eq:flux_1D}, this is the only ingredient needed to compute the total cross section, which in turn is crucial to extract the hidden particles yields in the detector. Up to this point, the simulation of a collimated (along the beam axis) but not mono-energetic beam of hidden particles particle impinging on the detector is achieved. However, this is not sufficient to develop a full event generator of unweighted ``interaction events''. A complete event reconstruction that gives access to 
correlations (between energies, positions, angles), may be of great importance. It can allow to study 
possible kinematical cuts with the aim of maximising the signal yield with respect to the backgrounds, and to accurately model detector effects. For example, due to different energy-angular correlations of the DM particles wrt the neutrino ones, it is possible to have a signal-enriched sample of events for off-axis detector configurations, as pointed out in ref.~\cite{Coloma:2015pih}. Hence, in principle this information might be useful to design optimised experimental configurations. Moreover, the output events can be further post-processed exploiting for example parton shower programs or other dedicated tools in order to have a better estimates of the detector efficiency.

\begin{figure}[h]
  \centering
  \includegraphics[scale=.3]{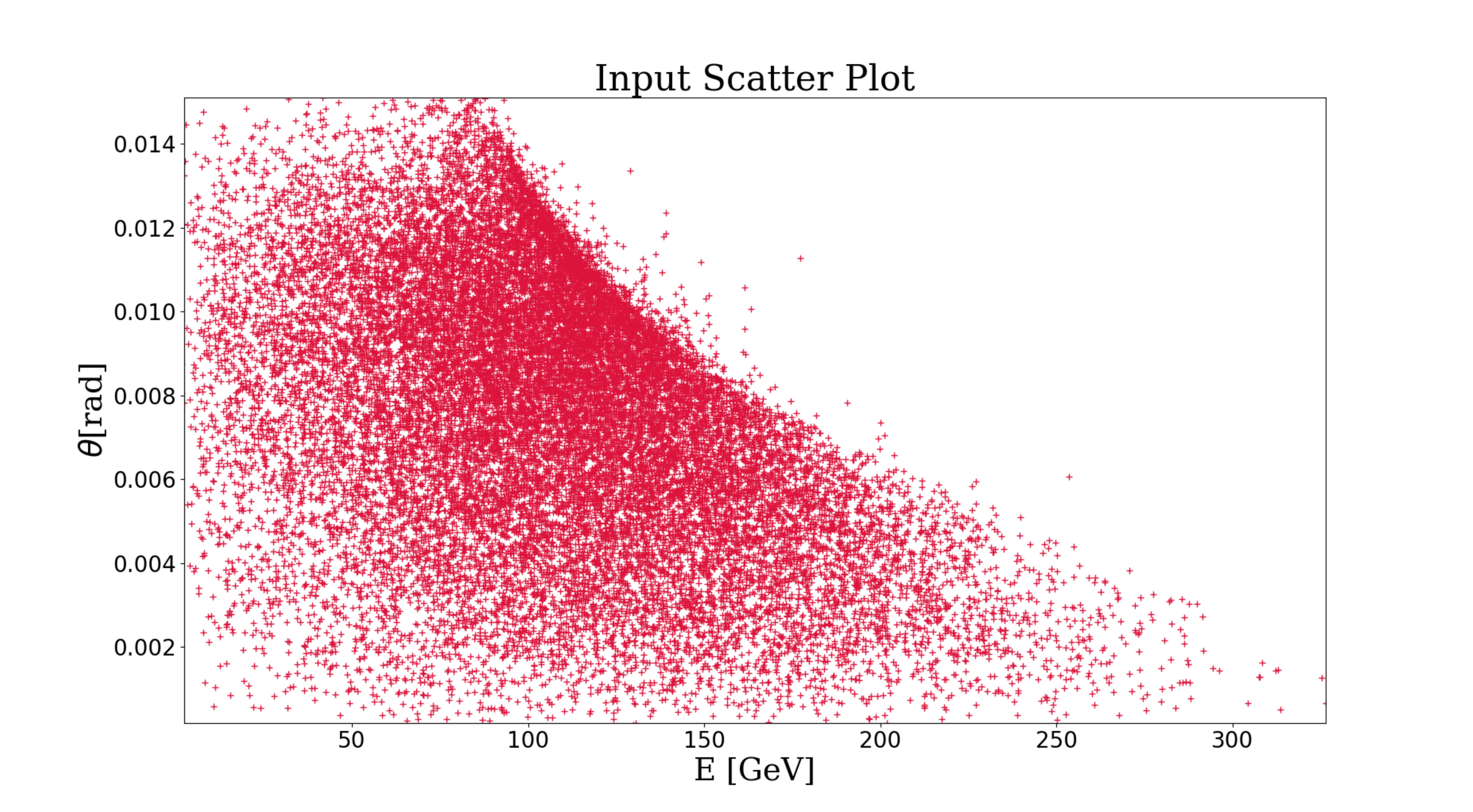}
  \caption{Input 2D-Scatter plot in the DM $E-\vartheta$ plan. As example, we consider DM particles produced by a Dark Photon-like mediator, which enter a SHiP-like detector}\label{fig::2d_fit_1}
\end{figure}

In principle, given the ``trivial'' dependence of the partonic cross section on the parameters $\vec{x}$, we 
can assign them a posteriori on a event-by-event base according to the distribution
\begin{equation}
  P(\vec{x})d\vec{x} = \phi(\overline{E},\vec{x})W(\overline{E},\vec{x}) d\vec{x}\,,
\end{equation} 
at fixed $E=\overline{E}$, where $\overline{E}$ is the energy of the current event. In practice, the procedure underlying the above formula is limited by the computational issue of performing a robust and reliable multi-dimensional fit, since the incoming particle flux is not known analytically. 
As mentioned above, our fitting procedure relies on the point-like approximation of the target in the primary interaction. 
\begin{figure}[h]
  \centering
  \includegraphics[scale=.75]{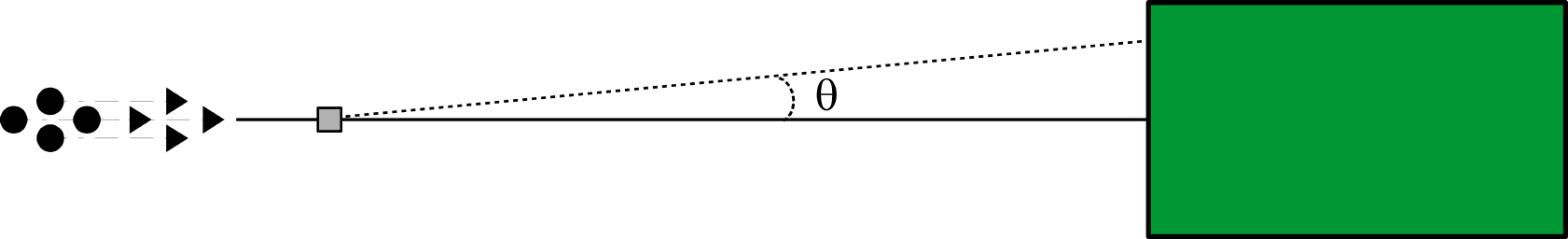}
  \caption{Production and detection of dark matter at a beam-dump experiment: a sketch of the setup and the kinematics.}\label{fig::setup}
\end{figure}
In a typical beam dump experiment, the distance between the production target and the near detector is greater than the characteristic size of the target, so that the point-like target approximation is a reasonable first approximation. Under this assumption, the complexity of the problem reduces considerably. As depicted in  Figure~\ref{fig::setup}, just three kinematical variables are needed to describe the incoming 
flux impinging the detector: the energy $E$, the polar angle $\theta$ and the azimuthal angle $\phi$. Furthermore, the physics occurring at the production point is invariant under a rotation around the beam axis resulting in flat distributions for the azimuthal correlations. Hence, the only relevant correlations are the $E-\theta$ ones. In Figure~\ref{fig::2d_fit_1} we show a typical plot of the production scatter data, which enter into the neutrino detector, in the $E-\theta$ plane for a SHiP-like configuration~\cite{Alekhin:2015byh,Anelli:2015pba}.

In order to generate $E,\theta$ values distributed in the same way, we have developed a numerical
2D-fitting algorithm which is fast, robust and automated. The main design concepts are based on the adaptive algorithms exploited in Monte Carlo integrators like VEGAS~\cite{Lepage:1980dq} and FOAM~\cite{Jadach:2002kn}. Our algorithm produces a 2D-mesh of bins for the 2D-histogram of the input points in the $(E,\theta)$-plane in such a way that the histogram heights are flat. It is based on a deterministic procedure: we apply a sequence of alternate splittings, one along the x-axis and one along the y-axis, according to a democratic principle of equal weights. Further details on this technique can be found in Appendix~\ref{Details1}. 
As an example, in Figure~\ref{fig::2d_fit_2}, we show the 2D-mesh associated to the scatter data in Figure~\ref{fig::2d_fit_1}.
\begin{figure}[h]
  \centering
  \includegraphics[scale=.3]{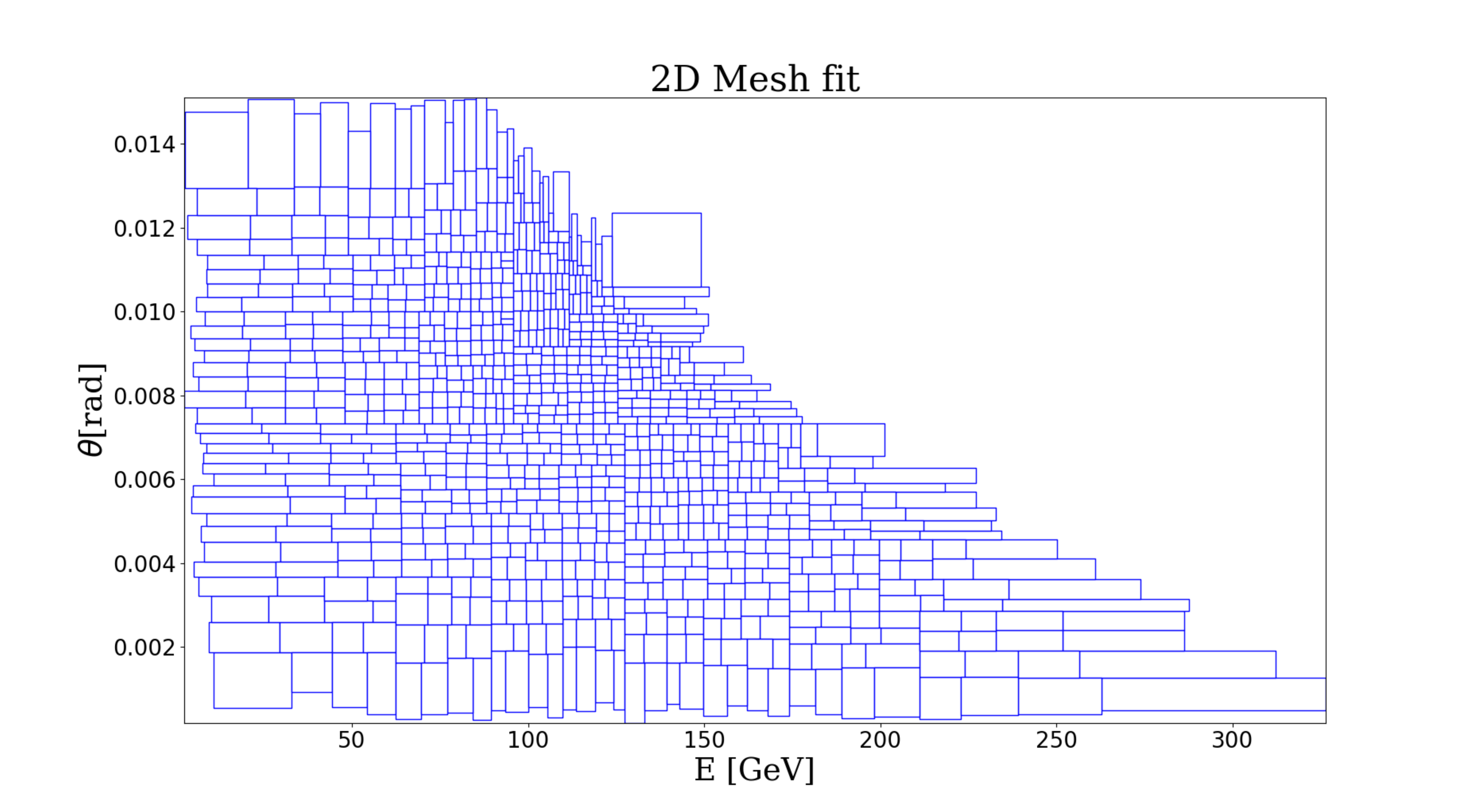} \\
  \caption{Result of the fitting procedure. The area of the cells in the mesh is proportional to the number of events in the cells.}\label{fig::2d_fit_2}
\end{figure}

\begin{figure}[h]
  \centering
  \includegraphics[scale=.3]{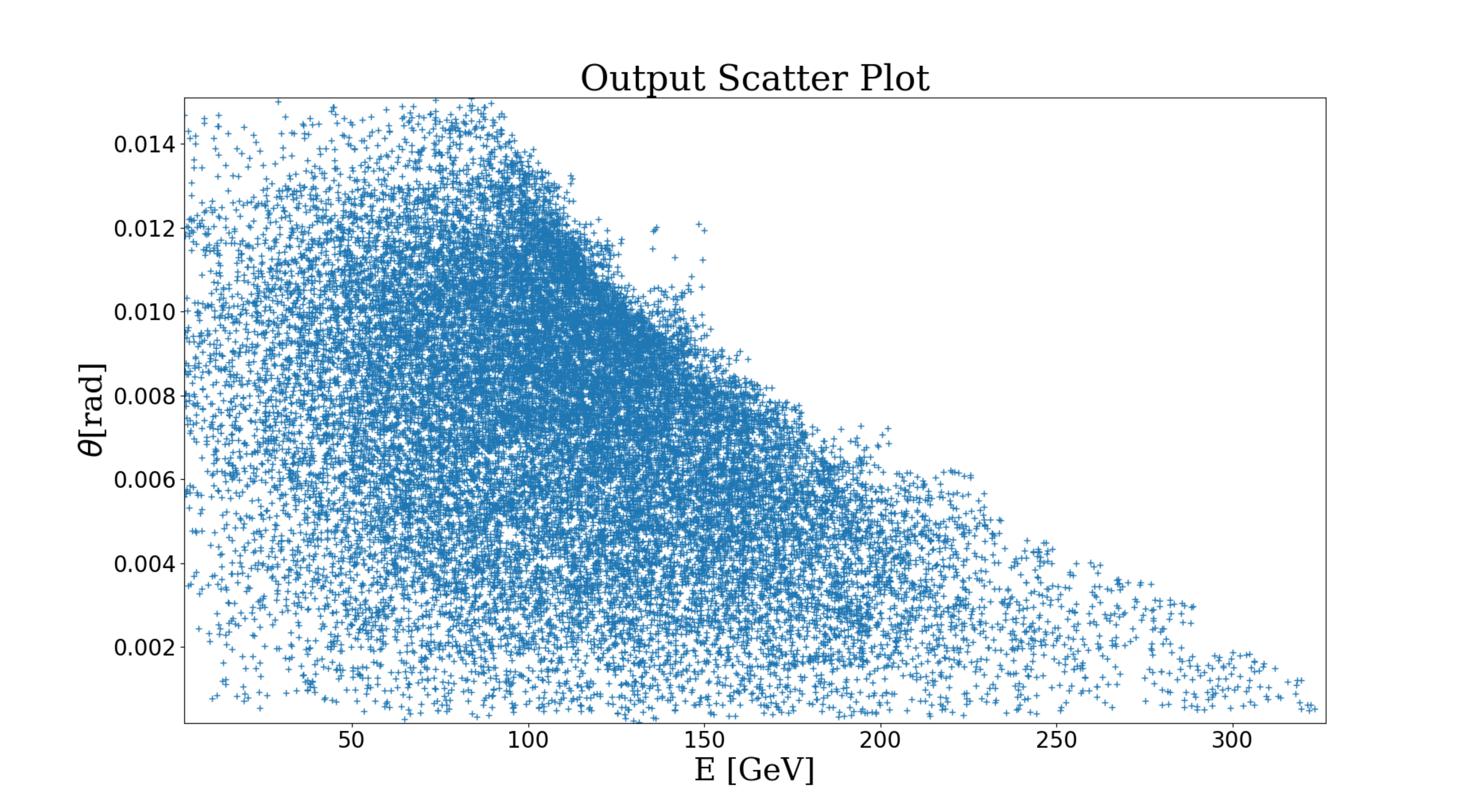}
  \caption{Re-generated 2D-scatter plot starting from the fitted 2D-mesh. See Figure~\ref{fig::2d_fit_1} for the comparison with the original plot.}\label{fig::2d_fit_3}
\end{figure}

Starting from this mesh, we can generate new $E,\theta$ points with the same distribution as the original ones. The result is plotted in Figure~\ref{fig::2d_fit_3}, where the goodness of the procedure can be appreciated by inspection.

\begin{figure}[h]
  \centering
  \includegraphics[scale=.3]{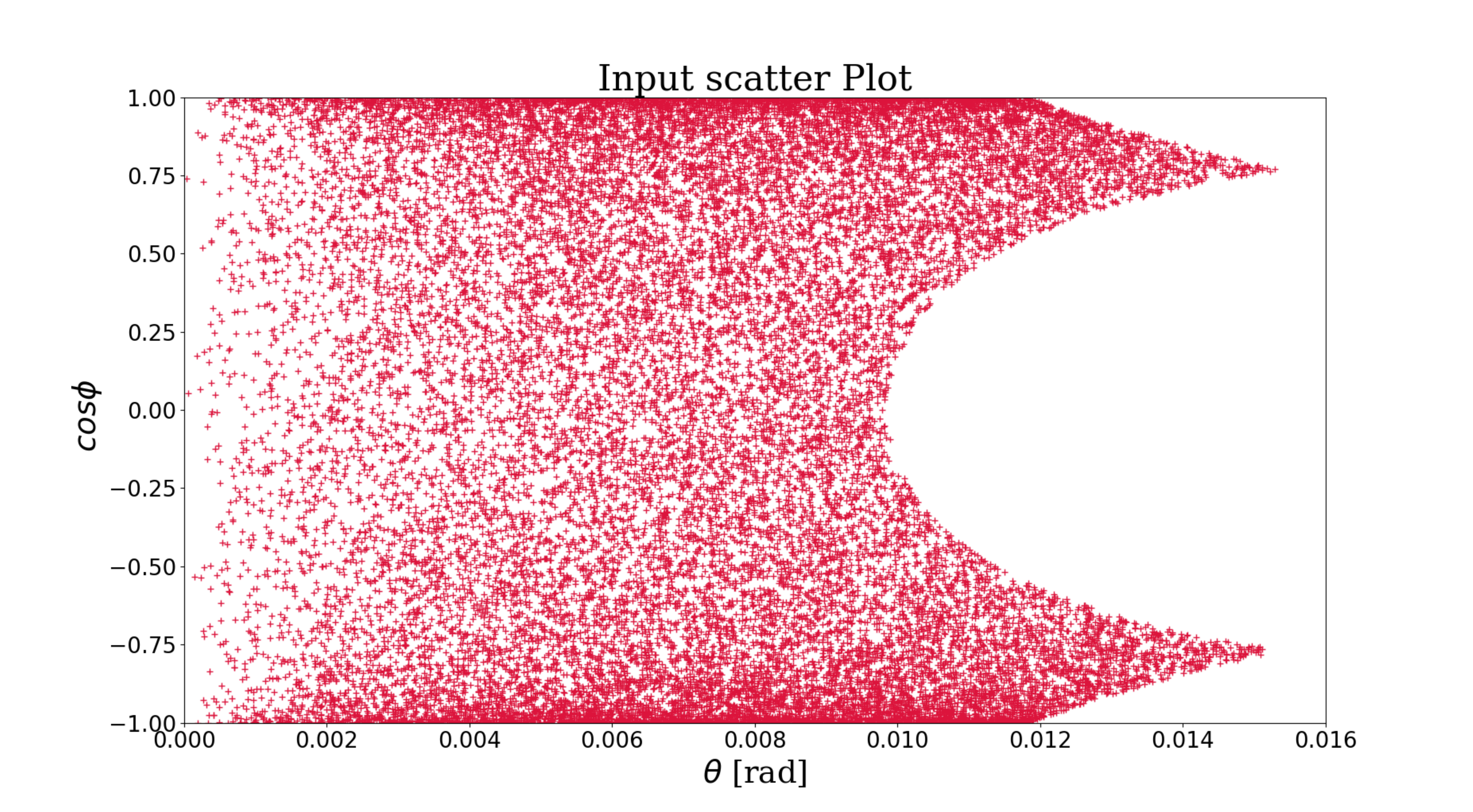} \\
  \includegraphics[scale=.3]{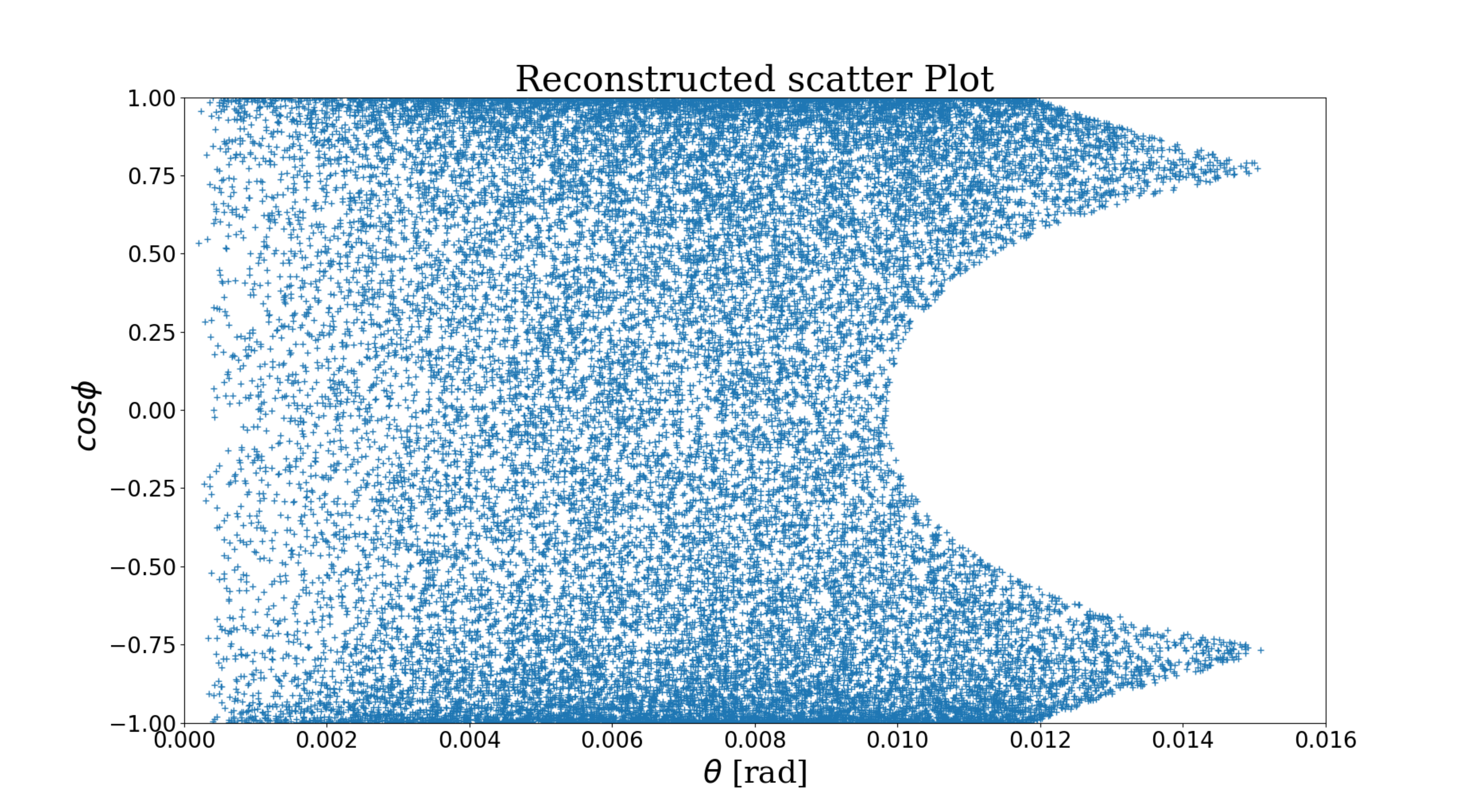}
  \caption{Comparison plot between the $\theta-\phi$ correlation as given by the original input scatter data (top panel) and by the re-generated points (bottom panel) obtained according to the procedure outlined in the main text.}\label{fig::azimuthal}
\end{figure}

We pass now to describe the generation of the azimuthal angle $\phi$. As already stated above, at fixed $\theta$, the $\phi$ distribution of the flux is flat due to the symmetry with respect to beam axis at the production point. The only kind of angular correlations which can be introduced are of geometrical nature and depend on the shape of the detector. An off-axis detector or even simply a detector on-axis with a rectangular surface exposed to the hidden particle flux, produces  $\theta-$dependent boundaries on $\phi$. To be definite let us consider the on-axis detector as in the SHiP experiment. In this case, at fixed $\theta$ value, the projection of the points at the detector surface obtained by varying the $\phi$ angle is simply a circle. If the circle
is entirely contained in the detector surface, one can generate the azimuthal angle flat in the full range $[0,2\pi)$. Otherwise, one must take into account the geometrical intersections between the circle and the surface, and generate a $\phi$ value flat only in the 'inside' region. In a simplified notation, our prescription states that we pick a $\phi$ value uniformly in the range
\begin{equation}
  \phi_\text{min}{\theta}\le \phi \le \phi_\text{max}{\theta},\quad \text{at fixed }\theta.
\end{equation}
In practice, it may happen that there are more than one interval for the azimuthal angle, as it is indeed the case for the rectangular detector we considered above. In that situation, for some values of $\theta$ there may be (it depends on the customary sizes of the sides) up to four intervals in which $\phi$ can lie, which correspond to the four corners of the rectangle. 
In Figure~\ref{fig::azimuthal}, we compare the angular correlations as obtained by the original scatter points which enter into the detector and the ones we reconstructed following our strategy. Again, the agreement is fairly good.
We stress that the above procedure is fully consistent with the re-weighting strategy outlined at the beginning, which we summarise here:
\begin{enumerate}
\item we first build the one-dimensional energy pdf on top of the original (unweighted) scatter points re-weighted by the effective function $W$, which takes into account, for example, that for some $\theta$ values, there are $\phi$ values not allowed;
\item we then reconstruct the missing variables in their actual ranges.
\end{enumerate}
We exploited the same re-weighting strategy to modeled the full 3D-geometry of the detector.
Indeed, up to this point, the same weight has been assigned to each particle direction. 
More technical details on this procedure are reported in Appendix~\ref{Details2}.
Depending on the specific new physics model, the interaction between hidden particles and the SM matter in the detector can be based on different types of processes:
\begin{itemize}
\item elastic scattering off electron;
\item DIS-like scattering off nuclei;
\item elastic and coherent scattering off nuclei. 
\end{itemize}
In principle,  the processes included in the above list can be easily simulated in our framework if a suitable model file is supplied, at least for the first two cases.

\subsubsection{Displaced decays}

In the case of displaced decays, the algorithm  simplifies considerably. The decay process does not require the regeneration of the events and then the issue of their full reconstruction does not arise.  Indeed, the decay can be generated event-by-event on top of the incoming flux of the unweighted events. The probability that a given particle decays in a specified decay channel $i$ after having traveled a distance $l$ from the production point is given by
\begin{equation}
  P_i(l) = \text{Br}_i \times (1 - e^{-{l}/{\lambda}}), \qquad \lambda= \frac{\gamma \beta}{\Gamma}
\end{equation} 
where $\text{Br}_{i}$ is the branching ratio for the $i$-channel, $\gamma$ and $\beta$ are, respectively, the Lorentz factor and the velocity in the laboratory frame of the decaying particle, and $\Gamma$ is its width. 
The displacement from the production to the decay point can be determined starting from the partial decay widths. The latter, for the given BSM model, are computed on-the-fly according to the actual parameters of the simulation ( the so-called ``auto-width" option). This feature in combination with the ``auto-scan" mode, both provided by \mg, allows for a complete simulation scan over the relevant parameter space. In the present version of {\sc MadDump}, the displaced decay events are not forced to be generated inside the actual decay vessel. It is only as the last step of the simulation,  that a rejection of the events that occur outside the detection volume happens. The algorithm could be improved by reweighting each event according to the distance that the decaying particle could actually travel inside the decay vessel.

\section{Illustrative examples}
\label{Examples}

In this section we provide some illustrative applications of {\sc MadDump}, considering three new physics
models and making the corresponding predictions for the SHiP experimental setup~\cite{Anelli:2015pba}. We stress that the SHiP facility configuration used in the following is based on the one developed for the Technical Proposal in 2015~\cite{Anelli:2015pba}. Since then, the SHiP Collaboration has continuously improved its setup aiming at higher sensitivity in the different channels. The newest setup as well as the corresponding background estimates are not yet available. The analyses reported in this paper will have to be redone once the updated information becomes available. Table~\ref{tab::SHiP_setup} summaries the relevant input parameters which specify also the geometry of the apparatus.
\begin{table}
\centering
\begin{tabular}{||c|c||}
	\hline
	parameter & value \\ [0.5ex]
    \hline
    \hline
    \# proton-on-target & $2\cdot 10^{20}$ \\
    \hline
    detector configuration & on-axis \\
    \hline
    distance target-detector & $5650\,$cm\\
    \hline
    detector density & $3.72$\,g/cm$^3$ \\
    \hline
	detector shape & parallelepiped \\
     & x-side: $187$\,cm \\
     & y-side: $69$\,cm \\
     & z-side: $200$\,cm \\
    \hline
   	detector efficiency & 1 \\
    \hline
\end{tabular}
\caption{Main input values used for the simulation of the SHiP detector geometry, as reported in ref.~\cite{Anelli:2015pba}.}
\label{tab::SHiP_setup}
\end{table}
In particular, we determine the number of detectable events for each
model in a typical run and then using the expected rates for the background processes reported in~\cite{Anelli:2015pba} we estimate the experimental sensitivities at $99.73 \%$ confidence level, which corresponds to the 3$\sigma$ contour,  and compare them with existing limits.

\subsection{Quark-DM scattering: leptophobic portals}

Let us first consider the case where an hidden particle that could be a dark matter candidate  interacts with the visible sector via a new leptophobic force. This is a good benchmark model to study quark-DM scattering, in particular we will focus on signatures of deep inelastic scattering.

\subsubsection{Vector portal: Baryonic $U(1)_B$  }

The simplest possibility for a leptophobic force mediated by a spin 1 particle is
provided by models where the baryon number $U(1)_B$ is gauged such as 
\begin{equation}
\mathcal{L}_{U(1)_B}=\mathcal{L}_q + \mathcal{L}_{\chi}+ 
\frac{m^2_{Z'} }{2}
Z^{\prime \mu} Z^{\prime}_{\mu}\, ,
\end{equation}
where the actual mass generation mechanism is not relevant here. 
The quarks are the only SM fermions charged under  this new gauge symmetry thus:
\begin{equation}
\mathcal{L}_q = \frac{g_Z}{2}  Z^{\prime}_\mu  \times \frac{1}{3} \sum_q  \; 
\overline q \gamma^\mu q \,, 
\label{eq:Lagrangian}
\end{equation}
while for the DM particle $\chi$
\begin{equation}
\mathcal{L}_\chi = \frac{g_Z}{2}  Z^{\prime \mu} \times \left\{ 
\begin{array}{c}  z_\chi \overline \psi_\chi \gamma_\mu \psi_\chi   \;  
 \\ [3mm] i z_\chi \left[ (\partial_\mu \phi_\chi^\dagger) \phi_\chi  - 
 \phi_\chi^\dagger  \partial_\mu \phi_\chi \right] 
 \end{array} \right. \,,
\end{equation}
where the only important requirement on $\psi_\chi $ is that it is long-lived enough to reach the detector.
The $U(1)_B$ is anomalous and the cancellation of anomalies could lead to 
additional strong constraints as discussed in~\cite{Dobrescu:2014fca,Dror:2017nsg} . 
However, these constraints depend on whether the anomalies are canceled by 
fermions chiral or not under SM gauge symmetries, hence they are UV-dependent 
and  we will not include them while comparing  sensitivity of various low energy probes. 
\newline
The existing bounds on the $Z'$ coupling in the 1--10 GeV mass range come from 
the $Z'$ exchange induced invisible decays of quarkonia  such as  $\Upsilon 
\rightarrow \chi \bar \chi $  and $ J/\psi \rightarrow  \chi \bar \chi) $ see~\cite{Graesser:2011vj}.
Monojet searches at hadron colliders set a bound on $g_Z$, the strongest one coming from a CDF search~\cite{Shoemaker:2011vi,Aaltonen:2012jb} at Tevatron, $  g_Z^2 \; {\rm BR}( Z' \to \chi \chi) < 1.4 \times 10^{-2}$. 
Moreover, existing  and previous neutrino facilities like MiniBooNE could have 
sensitivity to few GeV leptophobic $Z'$  as discussed in 
\cite{Dobrescu:2014ita,Coloma:2015pih,Frugiuele:2017zvx} where it is shown that a reanalysis of existing data 
could set the strongest bounds on an ample region of the parameter space.
\par In this study, we exploited the prompt production mode of {\sc MadDump} 
for the generation of dark matter particles in the s-channel for an 
almost on-shell $Z^{\prime}$. We use our own version for the
UFO file for this model,  which is available in the {\sc MadDump} directory.
Considering the relevant parameter, the mass of the $Z^{\prime}$, in the range 
$[2,10]\,$GeV with equal steps of $1\,$GeV, we have generated
$100\,$k production events (we remark that in each event the multiplicity of dark matter
particle is 2), while we generate $10$k DIS interaction events between the 
dark matter particles and the detector. The fraction of events which passes the
detector acceptance ranges from $\sim 14\%$ for $m_{Z^{\prime}}=2\,$GeV to 
$\sim 1\%$ for $m_{Z^{\prime}}=7\,$GeV. Exploiting a small workstation with sixteen cores, 
we have got the following timings for the complete simulation of each benchmark point:
\begin{itemize}
\item production : $\sim 1^\prime$
\item fit : $\sim 1^\prime$
\item interaction : $\sim 3^\prime$\,.
\end{itemize}
\begin{figure}[t!]
  \centering
  \includegraphics[scale=1]{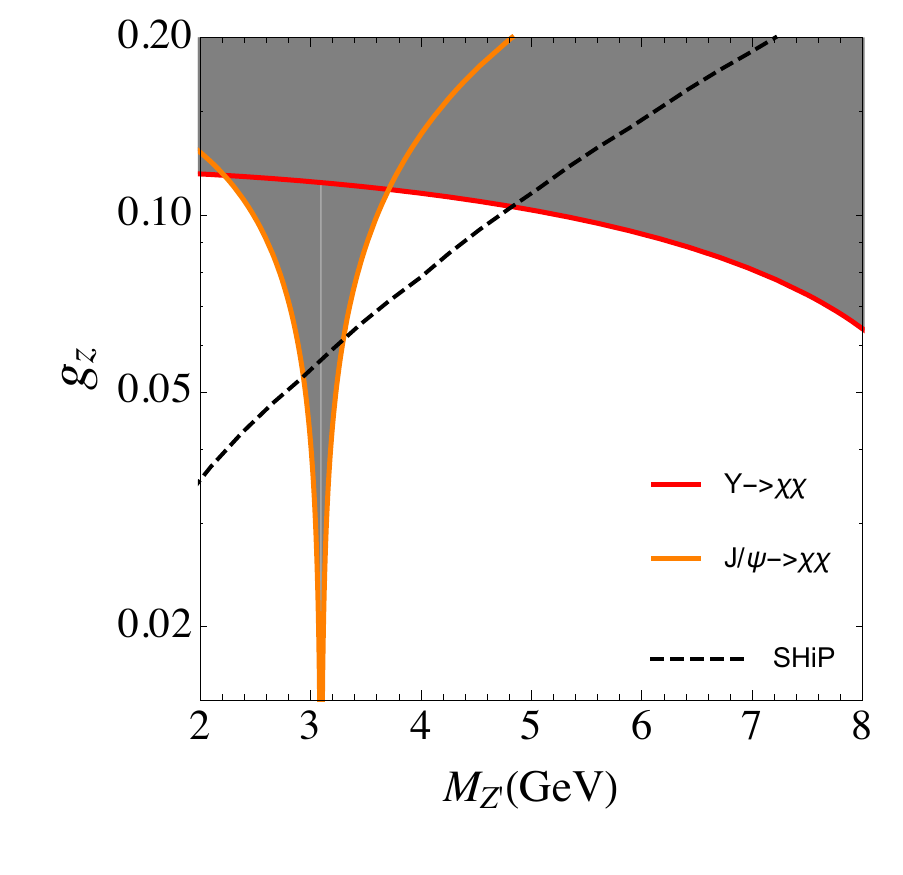}
  \caption{Sensitivity plot to a GeV leptophobic force at a SHiP like experiment obtained with
{\sc MadDump} assuming $ 9 \times 10^5$  neutrino deep inelastic background
events~\cite{Anelli:2015pba}. }
  \label{fig:LPforce}
\end{figure}
Based on~\cite{Anelli:2015pba} we assume $ 9 \times 10^5$  background events. In Fig.~\ref{fig:LPforce} we present the potential sensitivity of SHiP for $ 2 \times 10^{20}$ POT  to this scenario and we compare it with the above mentioned existing constraints.

\subsubsection{Leptophobic scalar and pseudo-scalar portal}

Another interesting possibility to consider is a leptophobic force mediated by 
a scalar or pseudo-scalar particle. 

We consider the following  simplified model 
\begin{equation}
\mathcal{L}_{S/a}=  \frac{ 1} {2}\partial^{\mu}{S}\partial_{\mu}{S} - \frac{1}{2} m_S^2 S^2 +S
\sum_i{  g_{S}^{ i}  \bar q_i q_i }- S \bar \chi \chi .
\end{equation}
with $i=u,d,s,c,b,t$; as before, $\chi $ is a Dirac fermion stable or long-lived enough to cross the SHiP detector.
In the case of a real scalar $S$ the interaction with quarks could arise via the renormalisable interaction
\begin{equation}
\mathcal{L}_{S} \supset  g_{S}S H^{\dagger} H\,,
\end{equation} which induces a singlet-Higgs mixing $\sin{\theta} $  such as the singlet $S$ inherits couplings to the SM fermions:
\begin{equation}
 g_{S}^{ i} = y_i \cdot \sin{\theta}   \,,
\end{equation}
where $ y_i$ is the SM Yukawa coupling of the fermion $i$. 
A different flavor structure from the SM for the singlet-fermion coupling  could be arranged via the dimension five operators, that is:
\begin{equation}
\mathcal{L}_{S} \supset  \sum_i{ ( \frac{ \tilde g_{S}^{ q_i q_i} }{\Lambda} 
S H_c \bar Q^i_{L } d^i_R} +{\rm h.c.})
\end{equation}
and
\begin{equation}
 g_{S}^{ i} =\frac{ \tilde g_{S}^{ q_i q_i} v }{\Lambda} \,,
\end{equation}
where $\Lambda $ is the cutoff above which either extra Higgs bosons or vector-like
leptons are expected. Depending on the origin of the interaction among  $S$ and the quarks, bounds from Higgs 
invisible decay and/or electroweak precision measurements could be relevant. However, 
we will  not discuss them due to their dependence on the UV-completion. 
Moreover, we assume that $CP$ is a good symmetry of the Lagrangian (see for 
instance~\cite{mckeenflavor} for a discussion of possible constraints).

We consider the benchmark scenario where the scalar $S$ couples to up and down quarks (see~\cite{mckeenflavor}). We have used the same setup of the previous case, $100\,$k prompt production events
and $10\,$k DIS interaction events. As for the UFO file for the scalar model, we exploited 
the general model in ref.~\cite{Mattelaer:2015haa}.
The program scanned over the relevant parameter, the mass of the scalar mediator $S$, 
in the range $[2,10]\,$GeV with equal steps of $1\,$GeV. 
The fraction of events which passes the detector acceptance ranges from 
$\sim 16\%$ for $m_{Z^{\prime}}=2\,$GeV to  $\sim 4\%$ for $m_{Z^{\prime}}=7\,$GeV.
The timings are analogous to those of the previous case.

\begin{figure}[h!]
  \centering
  \includegraphics[scale=1]{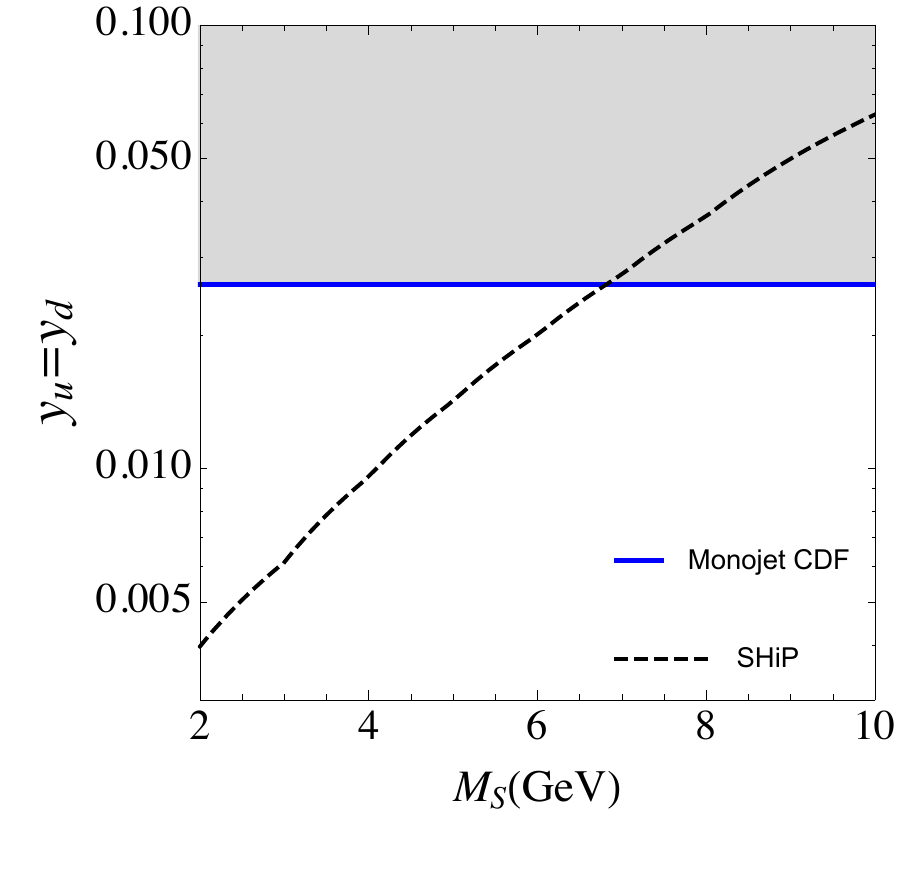}
  \caption{Sensitivity plot to a GeV scalar mediator at a SHiP like experiment 
  obtained with {\sc MadDump} assuming $ 9 \times 10^5$ neutrino DIS events  compared to existing bounds from CDF~\cite{Shoemaker:2011vi,Anelli:2015pba}.}
  \label{fig:GSmediator}
\end{figure}

In Fig.~\ref{fig:GSmediator} we show the sensitivity at SHiP considering as before DIS as signal events. In this case the only existing bounds come from the CDF monojets bounds~\cite{Shoemaker:2011vi,Anelli:2015pba} and we notice that SHiP could constrain new regions of the parameter space with this proposed analysis.

\subsection{Electron-DM scattering: the dark photon}
As benchmark model to study DM-electron scattering we consider a new 
gauge boson associated to an abelian gauge symmetry $U(1)'$,  $A'$, kinetically 
mixed with the photon~\cite{Holdom:1985ag}, namely a dark photon (DP).
The relevant Lagrangian corresponds to:
\begin{equation}
\mathcal{L}_{A^{\prime}} =- \frac{1}{4} F'_{\mu \nu}F^{\prime \mu \nu} 
+\frac{m^2_{A'} }{2}A^{\prime \mu} A^{\prime}_{ \mu}-\frac{1}{2} \epsilon \;  
F^{\prime}_{\mu \nu} F^{\mu \nu}\,,
\end{equation}
where $  \epsilon $ is the DP-photon kinetic mixing.
We further assume the existence of a particle $\chi $ either a scalar or a 
fermion charged under the new gauge symmetry $U(1)'$ and stable at least 
compared to the scale of SHiP, hence we also add the following Lagrangian:
\begin{equation}
\mathcal{L}_\chi = \frac{g_D}{2}  A^{\prime \mu} \times \left\{ 
\begin{array}{c}  \overline \psi_\chi \gamma_\mu \psi_\chi ,   \;  
 \\ [3mm] i \left[ (\partial_\mu \phi_\chi^\dagger) \phi_\chi  - 
 \phi_\chi^\dagger  \partial_\mu \phi_\chi \right] 
 \; ,  
 \end{array} \right. \,.
\end{equation}
We choose as benchmark point $ m_{\chi}= m_{A'}/3 $ and $\alpha_D = 
\frac{g_D^2}{4 \pi}=0.5 $ as in~\cite{Battaglieri:2017aum}.  

\par
\begin{figure}[h]
  \centering
  \includegraphics[scale=.6]{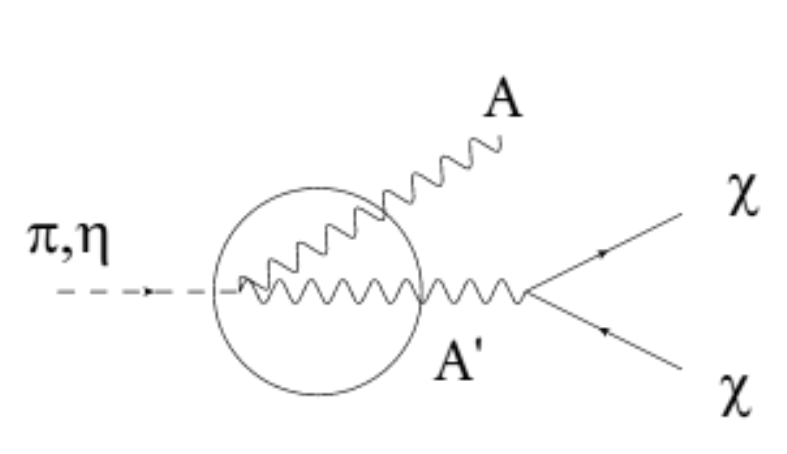}
  \caption{Effective field theory approach to production of DP from rare meson decay. }
  \label{fig:eft_vertex}
\end{figure}

 For this case study, we have used the decay-interaction mode of {\sc MadDump}. 
The incoming meson fluxes has been generated with {\sc Pythia8}, having care to store
only the final state mesons which decayed directly in photons. This means that in
a decay chain $\eta \to 3\pi_0 \to 6\gamma$ only the $3\pi_0$ are stored, while 
a $\eta$ meson is stored in the list if it decayed directly into $2\gamma$.
With this caveat, we can limit ourselves to consider only one decay channel 
for each of the mesons included in our analysis:  
\begin{itemize}
\item $\pi_0 \to 2\gamma$;
\item $\eta \to 2\gamma $; 
\item $\omega \to \gamma\pi_0 $.
\end{itemize}
We exploited the general UFO model for spin-1 as reference model for the DP~\cite{Mattelaer:2015haa,Backovic:2015soa}.
The meson decays has been modeled applying a standard effective 
field theory (EFT) approach. Indeed, for the most interesting case in which 
the DP is almost on shell, the decay process can be approximated
by the tree-level vertex depicted in Fig.~\ref{fig:eft_vertex} at first order
in the EFT expansion. 
We added the meson particles and the minimum set of new interactions required to deal 
with their decays directly in the UFO file, on top of the reference model.
We simulated $100\,$k proton on target (POT) events, which, in terms of meson yields/POT, 
resulted in: $\sim 6/$POT for pions, $\sim 0.3/$POT for $\eta$'s and $\sim 0.07/$POT
and for $\omega$'s. 
We considered one meson species at time, and scanned over the relevant parameter space, for
masses of the DP below the corresponding meson mass. For the case of the pions, which are the most
numerous particles, the time per scan point has been $\sim 20^\prime$ on a 4-cores CPU.
The most time consuming tasks are the I/O operations related to the meson decay process, 
which took $\sim14^{\prime}$ of the whole time per benchmark point.  

\begin{figure}[t!]
  \centering
  \includegraphics[scale=1]{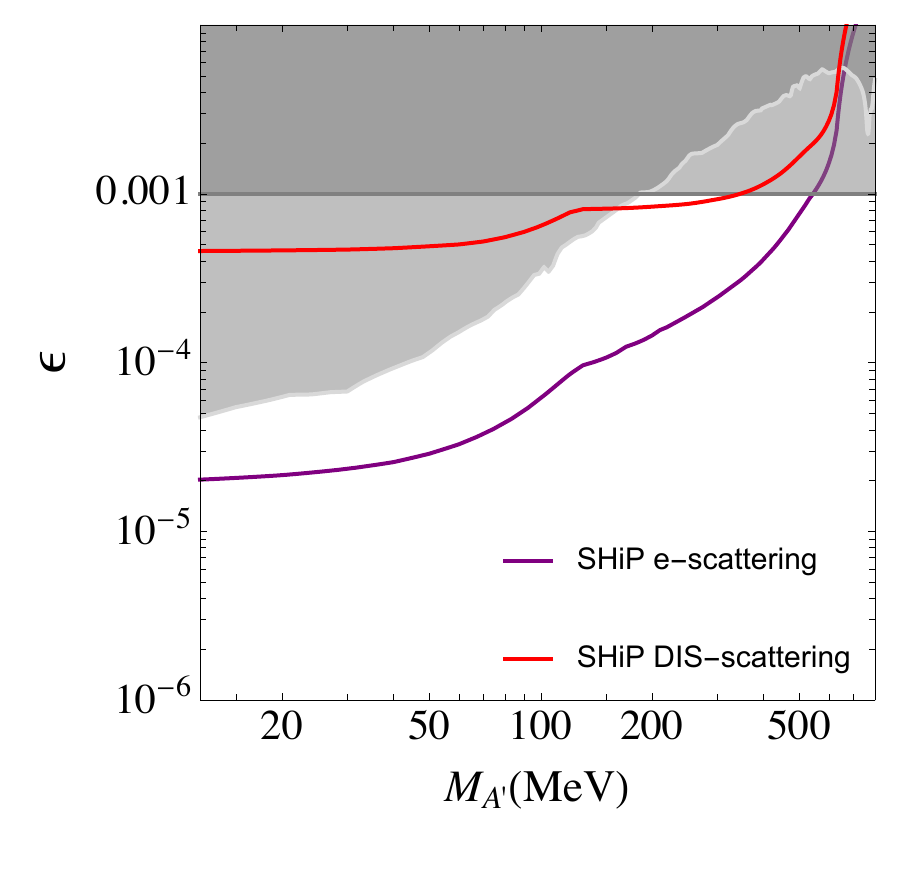}
  \caption{SHiP  sensitivity for a Sub-GeV dark photon emitted by rare meson decays of $\pi^0,\eta$ and $\omega$ obtained with {\sc MadDump}. For all production modes both electron-DM scattering and DIS are considered considering respectively 284 and 9000 background events. The gray region is excluded by current probes, the strongest are BaBar mono-photon search~\cite{Lees:2017lec} and MiniBooNE~\cite{minibooneE}. The dark gauge couplings is chosen to be  $ \alpha_D =0.5$ and the ration between DP and DM mass is 3.  }
  \label{fig:SGdp2}
\end{figure}

In Fig.~\ref{fig:SGdp2} we  present  in the $ (m_{A'}, \epsilon) $ plane the SHiP sensitivity compared to existing  bounds, described in details 
in~\cite{Battaglieri:2017aum}.  In  the region of interest, the strong
experimental constraints come from the monophoton BaBar search~\cite{Lees:2017lec} and
NA64~\cite{Banerjee:2017hhz} via a missing energy analysis. Assuming 
$ \alpha_D =0.5$, experiments looking at electron-DM scattering such as MiniBooNE~\cite{minibooneE}, LSND~\cite{lsndpatrick}, and E137~\cite{Batell:2014mga} achieve a better sensitivity than NA64 so their reach is also presented here. An even stronger reach for $ m_{A'} \lesssim 300$ MeV could be reached by NO$\nu$A experiment at Fermilab by recasting existing data as shown in~\cite{deNiverville:2018dbu}.
For electron scattering events, according to the SHiP technical proposal~\cite{Anelli:2015pba}, we consider
the following selection cuts: 
\begin{itemize}
\item $1\,\text{GeV}\le E_e\le 20\,\text{GeV}$
\item $10\,\text{mrad} \le \theta_{\chi-e} \le 20\,\text{mrad}$
\end{itemize}
where $\theta_{\chi-e}$ is the angle between the incoming DM particle and the outgoing electron.
We assume 284 electron-neutrino scattering 
events as background following~\cite{Anelli:2015pba}. 
We simulated events both for electron-$\phi_\chi$ scattering and DIS events comparing their sensitivity in Fig.~\ref{fig:SGdp2}.
As expected, the electron sensitivity is significantly better than the one achievable with DIS.
Our prediction here is conservative because we do not include potentially important contributions to the production stage like the decays from mesons produced in the cascade process and the prompt production.

\section{Conclusion}
\label{Conclusions}
In this paper we have presented a new {\mg} plugin called {\sc MadDump}
that allows the generation of events where the production of a particle and 
its detection are separated by a long distance. In order to install it, 
it is enough to type ``install maddump'' within \mg\footnote{For further details, please refer to \url{https://launchpad.net/maddump}}.
The main input provided by the user are the geometry of the experiment
and the physics model under investigation. With these ingredients
event generation corresponding to one or more benchmark scenarios can be performed automatically.  
We have shown illustrative examples based on different BSM scenarios and production/detection mechanisms and computed the corresponding number of events that would be produced at the SHiP experiment.
The framework is fully general and can be applied to any BSM model 
and experiment at a beam dump facility that aims to test it.  Our tool could be employed for a number of studies, from the search of new feebly interacting particles to the study  of elusive SM processes like tau neutrino cross section at present and future beam dump experiments. 

\section*{Acknowledgments}
We are thankful to Eduardo Cortina, Giovanni De Lellis, Antonia Di Crescenzo, Jan Hajer and Roberta Volpe for useful comments on the manuscript.
LB thanks Zahra Ghorbani Moghaddam for useful conversations. 
The work of LB and FT has been supported in part by the Italian Ministry
of Education and Research MIUR, under project n$^{\rm o}$ 2015P5SBHT
and by the INFN Iniziativa Specifica ENP. FM is supported by the 
European Union's Horizon 2020 research and innovation programme
as part of the Marie Sklodowska-Curie Innovative Training Network MCnetITN3 (grant
agreement no. 722104) and by the F.R.S.-FNRS under the `Excellence of Science` EOS
be.h project n. 30820817.

\appendix
\section{Techniques for event generation}
\label{Details1}
Consider the problem of generating unweighted points in a 2D-space according
to the distribution
\begin{equation}
  P (x, y) dxdy = f (x, y) g (x, y) dx dy\,,
\end{equation}
where $g (x, y)$ is a modulation function whose expression is supposed to be
known analytically. More precisely, we are mainly interested in the problem of
generating unweighted $y$ values given a fixed $x = \overline{x}$ according to
the profile function
\begin{equation}
\label{eq::profile}
  \phi (y) dy = P \left( \overline{x}, y \right) dy\,.
\end{equation}
When the function $f (x, y)$ is given in closed form the problem reduces to generating points according to a given function $P (x, y)$
and it can be accomplished by standard Monte Carlo techniques, using for example the classic
hit-or-miss algorithm. Here we consider the more interesting situation in
which the function $f (x, y)$ is only available numerically indirectly from sample of events. 

One can re-interpret it as a fitting problem. The function $f (x, y)$ is given in an
approximated way, with a level of precision which can be in principle reduced
at will (by generating more points) but at each step it is finite, as a
2D histogram built out of events. In its standard formulation, the task of obtaining a fit from an histogram consists basically of two different parts:
\begin{itemize}
  \item the choice of the model, i.e. an n-parameter family of functions together with
  the cost function of the fit;
  \item minimisation of the cost function. 
\end{itemize}
In so doing the result is given by a function supplemented with extra
information on the accuracy of the fit (covariance matrices, goodness of the
fit, etc). Fitting a function, even in the ``simple'' 2D case, is however not 
always trivial. In particular, aside the technical aspects
underlying the minimisation procedure, a certain amount of knowledge of the
function to be fitted is required (in order to choose a reasonable class of
models). For our purposes, no \textit{apriori} assumptions can be formulated
on the behavior of the function $f (x, y)$, as, in general, it can result from very 
different classes of physical processes. For this reason, we look for a procedure that allows to automatize the process. 

Though an analytical fit has advantages (including also the possibility of  smoothing a discrete data sample in to
continuous distribution), such a level of accuracy is not strictly required in order
to perform the generation of the unweighted events and we do not adopt it. Our approach is based on importance sampling and variance reduction methods implemented in Monte Carlo
integrator algorithms. The strategy is based on the flattening of the integrand function via a numerical adaptation of the integration grid.
Moreover, once the grids are available, they can be used to regenerate unweighted points according to the integral function.

In our implementation, we have devised a complete deterministic procedure to construct a grid, very closely the above concept of adaptive grid. In what
follows, we will give a detailed description of our algorithm together with some validation examples.

\subsection{Grid construction}

As a first example, let us consider the case in which the modulation function $g (x, y)$ reduces to the identity map. As simple as 
it may appear (in this case $P (x, y) dxdy = f (x, y) dxdy$ and  an unweighted generator for that distribution is assumed to be known), it allows us to clarify a few useful points.  First, it may happen that generating events
with the grid is more efficient (for what concern both time and space
resources), or more usable in some sense, than exploiting the original
generator. This is in fact the case in our applications, in which the
unweighted generator has a very complex structure and the 2D events we interested in are a tiny part of the whole result. For this
reason, it is not only an illustrative case, leading to a clearer illustration
of the basic concepts, but it is relevant \textit{per se}.

We assume to have at our disposal a sample of $N$ unweighted points $(x, y)$ distributed according to the function $f (x, y)$. Our aim is to generate unweighted events
distributed according to the same distribution. Starting from the available
points, the profile of the function $f (x, y)$ is given by the heights of a
2D histogram with bins of equal size. The idea is that of resizing the bins in such a way to flatten the histogram, or, equivalently,
to have the same number of points lying in each bin. In this way, the
distribution of the bins will follow the behavior of the function: they will
be denser and smaller near the region where $f (x, y)$ is peaked and sparse
and bigger where it is flat. The resulting 2D map will retain almost the full
information of the 3D plot, and it is very similar to the idea of a contour
plot.

In order to obtain such a parametrisation, we employ  a decision-tree-like algorithm, which is very simple and efficient. Before describing it, a technical remark is needed. Adaptive
grids are often constructed with lines parallel to the axis coordinates. This
is efficient in all the situations where the function can be expressed in the
factorised form
\[ f (x, y) = f_1 (x) f_2 (y) . \]
A great improvement is given by an approach in which irregular grids, made of
cells of different sizes, are allowed, as in the case of the FOAM algorithm~\cite{Jadach:2002kn}.
The cells adapt better to the behavior of the function reproducing it in a
more faithful way, for example near circular peaks. The cell represents the basic object of our algorithm. A cell can be split in two cells along the $x$-direction (horizontally) or
the $y$-direction (vertically). Given these basic ingredients, the algorithm
proceeds as follows:
\begin{enumerate}
  \item start from a cell containing all the available points;
  \item alternate an \ horizontal split and a vertical split, in such a way
  that, in each of the two splits, half of the point fall in a subcell and
  half in the other one;  
  \item repeat step 2 for each subcell until the number of point for cell
  is lesser/equal than a prefixed value (\textit{exit condition parameter}).
\end{enumerate}
It is clear that the above procedure gives the grid we are looking for. The
exit condition parameter $n_{min}$ controls the grain of the mesh. The
choice of its value is based on the compromise between having it small for a
finer grain and having a sufficient number of points per bin to be
statistical significant.

We now restore the proper role of the modulation function $g (x, y)$, which
as mentioned above, can be arbitrary yet to be expressed in an analytical form. It
can be viewed as a reweighting of the original sample of points:
\[ (x_i, y_i, 1) \rightarrow (x_{i,} y_i, g (x_i, y_i))\,, \]
where we have conventionally set to $1$ the common weight of the unweighted
sample. We have
\[ \langle P \rangle_{\rm uniform} = \int g (x, y) f (x, y) dxdy = \int g (\xi,
   \eta) d \xi d \eta = \langle g \rangle_{\rm f} \]
where the notation $\langle \cdot \rangle_{\rm pdf}$ denotes the average wrt the pdf in
the subscript. Under the hypothesis $f (x, y)$ is a distribution function, a
well-defined change of variables is implicitly given by the relation
\[ f (x, y) dxdy = d \xi d \eta \]
where the function $f$ is the Jacobian of the transformation. Furthermore, if
also $g$ has a definite sign it is possible to perform an extra
change of variables
\[ \langle g \rangle_{\rm f} = \int g (\xi, \eta) d \xi d \eta = \int dsdt = \langle 1 \rangle_{\rm f g}
   . \]
This relation proves the equivalence between
unweighted generation of the product distribution $f \cdot g$ and the
generation reweighted by $g$ starting from a sample of unweighted points
generated according to $f$.

We are now ready to generalise the previous case. We require that the rebinning procedure
leads to a grid with (almost) the same weight $w$ for each bin, with the
definition of the weight $w (b)$ of the bin $b$ given by
\[ w (b) = \sum_{(x_i, y_i) \in b} g (x_i, y_i) . \]
The generalisation of the algorithm is straightforward
\begin{enumerate}
  \item start from a cell containing all the available points;
  \item alternate an horizontal split and a vertical split, in such a way
  that, in each of the two splits, each subcell have half the weight;
  
  \item repeat step 2 for each subcell until its weight is greater than
  the prefixed value $w$ (\textit{exit condition parameter}).
\end{enumerate}
The exit condition parameter can be  chosen of the form
\[ w = \alpha \frac{w_{tot}}{N} \times n_{\min} \]
where $n_{\min}$ has the same meaning as before and $\alpha$ is a dimensional
factor which can be adjusted in the direction of refining the grain or
increasing the number of points per bin.
\par By construction, the above procedure cannot handle distributions
which vanish on some regions inside the fitted domain. This limitation
is particularly severe in the case the distribution presents a falling-down tail
and vanishes inside the fitted region. Indeed, even if the cells become larger and larger
when approaching the tail, there is always a non null probability to generate
points inside them, also in the empty regions. In this way, unphysical points
are generated. In order to milder this limitation, we implemented a further refinement step 
after the mesh has been constructed. The peripheral cells, i.e. the cells which share
a side with the frame of the fitted regions, are reshaped in a such a way to limit
the cell to the actual region populated by the input points. We refer again to Fig.~\ref{fig::2d_fit_2}
in section 2 to appreciate the reliability of this improvement. 

\subsection{Example}

\begin{figure}[h]
  \centering
  \begin{subfigure}[t]{0.5\textwidth}
    \centering
    \includegraphics[scale=0.25]{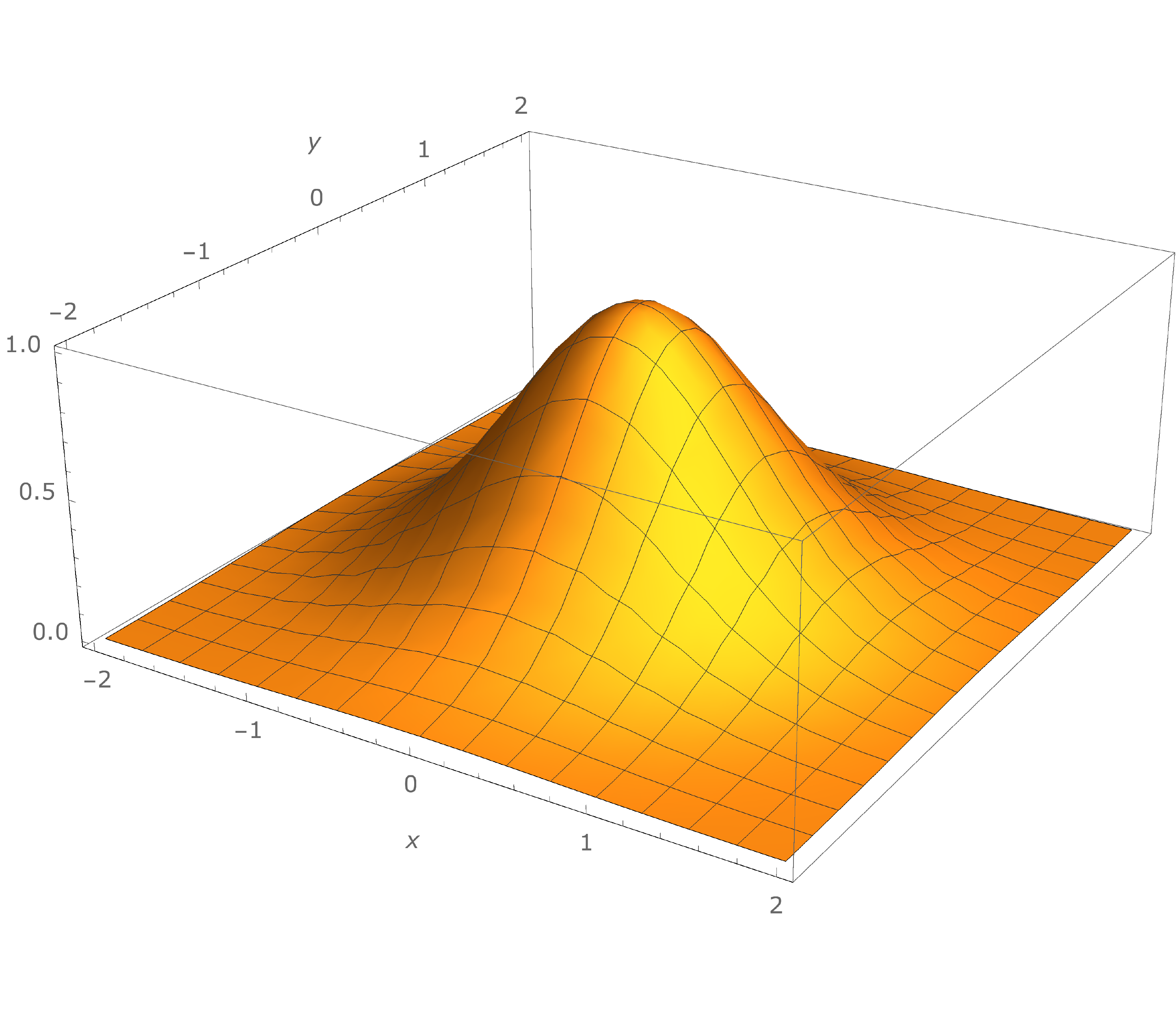}
    \caption{\footnotesize{Simple gaussian distribution as f function.}}
    \label{fig::ex_1_a}
  \end{subfigure}~
  \begin{subfigure}[t]{0.5\textwidth}
    \centering
    \includegraphics[scale=0.25]{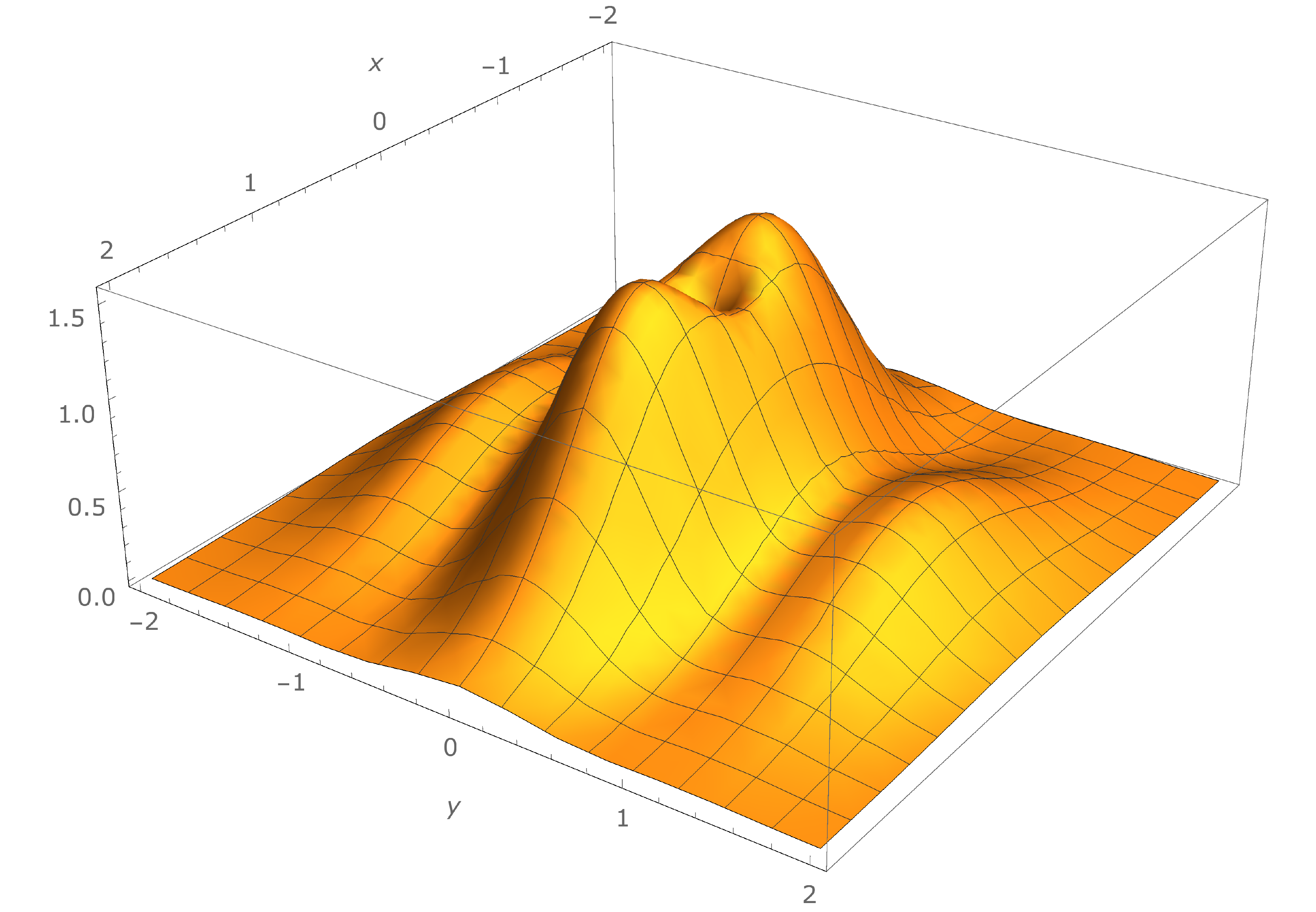}
    \caption{\footnotesize{Full distribution after modulation is applied.}}
    \label{fig::ex_1_b}
  \end{subfigure}
  \caption{3D-plot of the analytical distributions occuring in the validation example.}\label{fig::ex_1}
\end{figure}

As a validation example, we consider the situation in which both the function $f (x,y) $ and the modulation $g (x, y)$ are given analytically in order to show that the  algorithm works correctly. Furthermore, we test its robustness considering
a highly non-trivial case in which the modulation affects and distorts in a
severe manner the original function. We take a simple
Gaussian function, (see Fig.~\ref{fig::ex_1_a}):
\[ f (x, y) = e^{- (x^2 + y^2)} \]
with the following modulation
\[ g (x, y) = 2\sqrt{x^2 + y^2}\cos^2 (2 y) + 1 . \]
The product function is shown in the 3D plot in Fig.~\ref{fig::ex_1_b}. 

\begin{figure}[h]
  \centering
  \begin{subfigure}[t]{0.9\textwidth}
    \centering
    \includegraphics[width=1\linewidth,scale=.3]{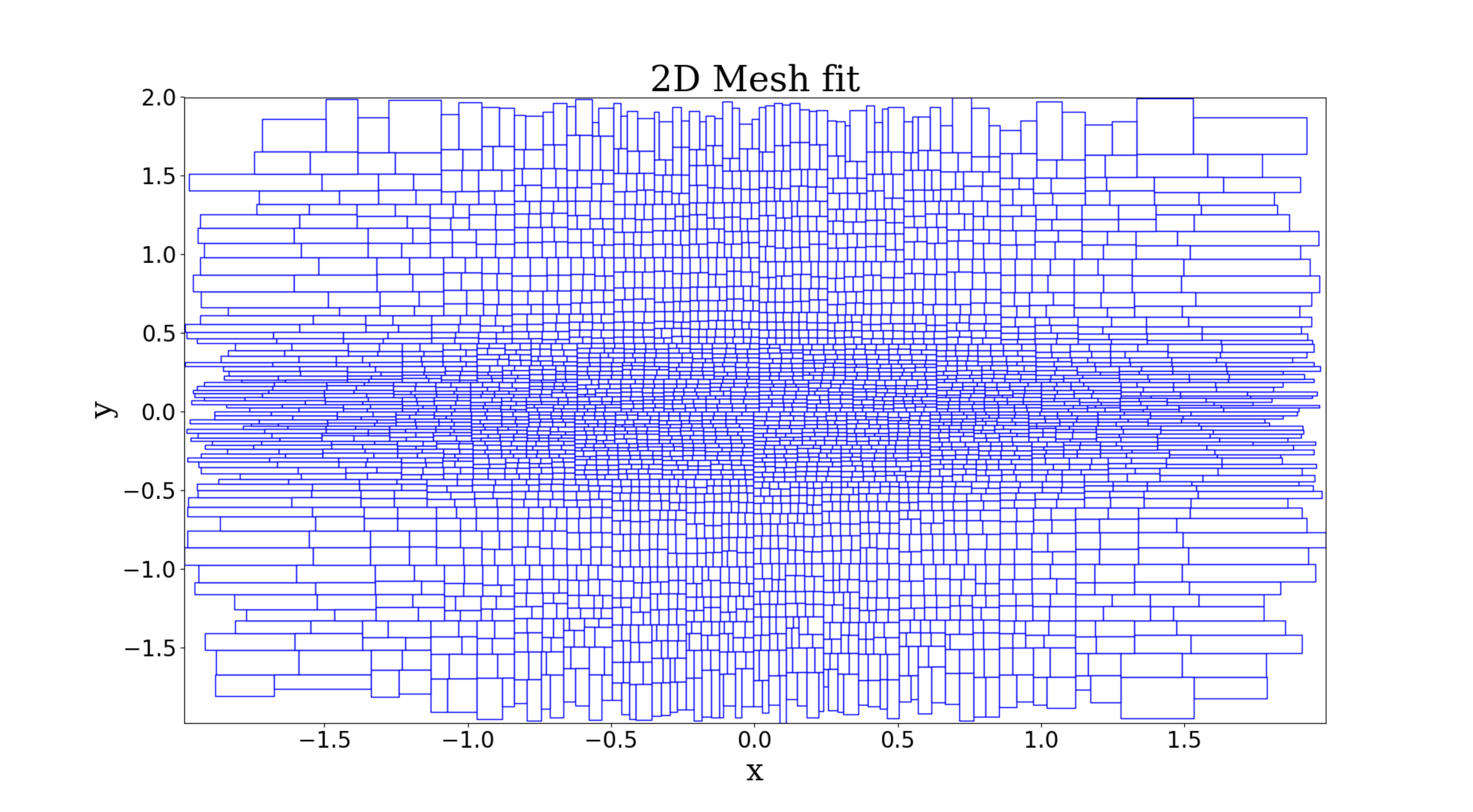}
    \caption{\footnotesize{$100\,$k input points}}
    \label{fig::ex_mesh_a}
  \end{subfigure}
  
  \begin{subfigure}[t]{0.9\textwidth}
    \centering
    \includegraphics[width=1\linewidth,scale=.3]{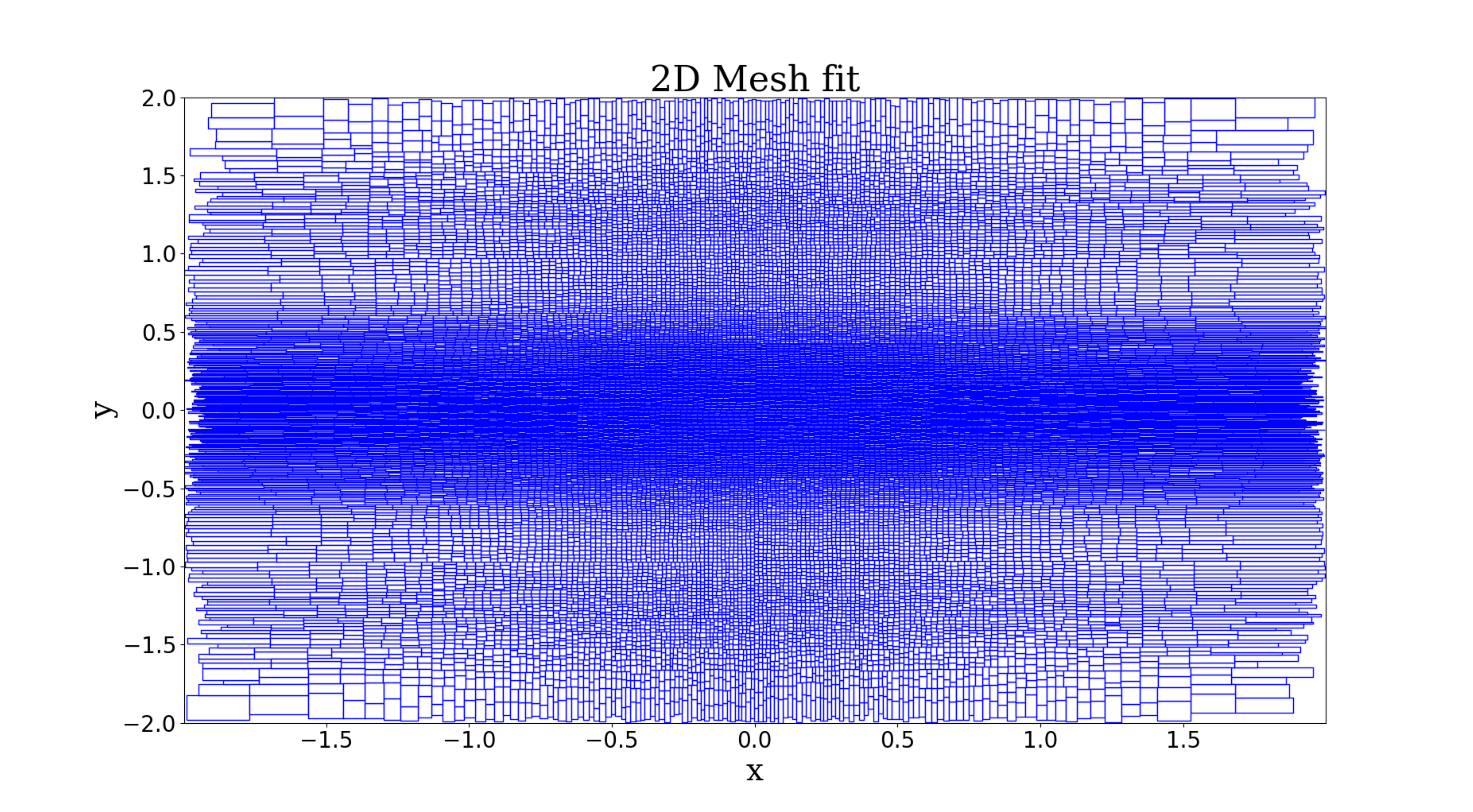}{b}
    \caption{\footnotesize{$1\,$M input points}}
    \label{fig::ex_mesh_b}
  \end{subfigure}
  \caption{2D mesh obtained with our algorithm for increasing number of input points.}\label{fig::ex_mesh}
\end{figure}

In Fig.~\ref{fig::ex_mesh_a} and Fig.~\ref{fig::ex_mesh_b} we report the corresponding 2D meshes
obtained with our algorithm for $100\,$k and $1\,$M input points respectively.  
We stress that the starting point has been the generation of a sample 
of unweighted points distributed according to the Gaussian function. 
We have reweighted the points according to the modulation function and then 
we have applied our algorithm for weighted events. The algorithm reproduces faithfully the behaviour of the  function with a level of accuracy which, as expected,  improves with the number of input points. 

\subsection{Generation of unweighted events}

Generating a 2D sample starting from the 2D mesh with the same distribution as $f(x,y)$ is trivial. By construction
\begin{itemize}
  \item the probability of generating a point in a given cell is proportional
  to the inverse of its area,
  \item inside a cell, the probability of generating a point is uniform, \ 
\end{itemize}
and therefore it is enough to generate an equal number of points uniformly in each cell.

Let us now turn to the issue of generating an $y$ value according to the profile
function Eq.~\eqref{eq::profile} at a given $x = \overline{x}$ point. We introduce a small
resolution parameter related to the $x$ variable $\epsilon$ such that the $x =
\overline{x}$ value is fixed within the interval $\left[ \overline{x} -
\epsilon, \overline{x} + \epsilon \right]$. Then, the thin stripe centered in
$x = \overline{x}$ with width $2 \epsilon$ and parallel to the $y$-axis will
intercept the mesh in a subset $S$ of cells.

We associate a normalised weight to each cell
in $S$ proportional to the ratio of the overlapping area between the stripe
and the cell over the total area of the cell. Then, we pick a cell according
to the value of these weights by generating a uniform random number in the
interval $[0, 1]$. Finally, we generate a uniform $y$ value within the cell.
This construction solves our problem, i.e. the $y$ values are distributed according to the profile function $\phi$. The procedure
is independent of $\epsilon$ in the limit $\epsilon \rightarrow 0$. In
practice, this means that magnitude of $\epsilon$ should be chosen as a
fraction of the minimum $x$-width of the cells of the mesh. 

\subsection{Example}

\begin{figure}[h]
  \centering
  \begin{subfigure}[t]{0.5\textwidth}
    \centering
    \includegraphics[scale=0.18]{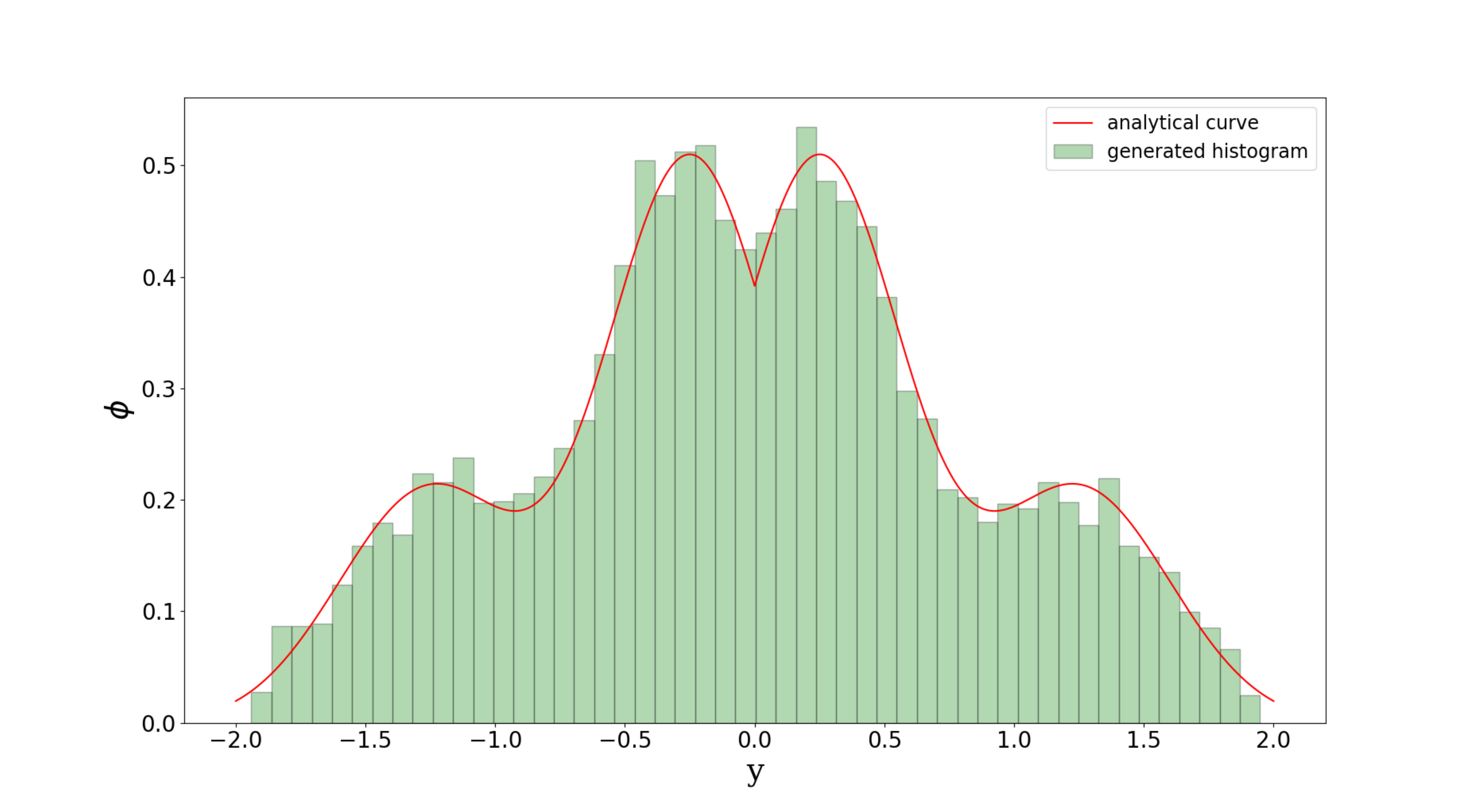}
    \caption{\footnotesize{Using the $100\,$k mesh.}}
    \label{fig::ex_regen_a}
  \end{subfigure}~
  \begin{subfigure}[t]{0.5\textwidth}
    \centering
    \includegraphics[scale=0.18]{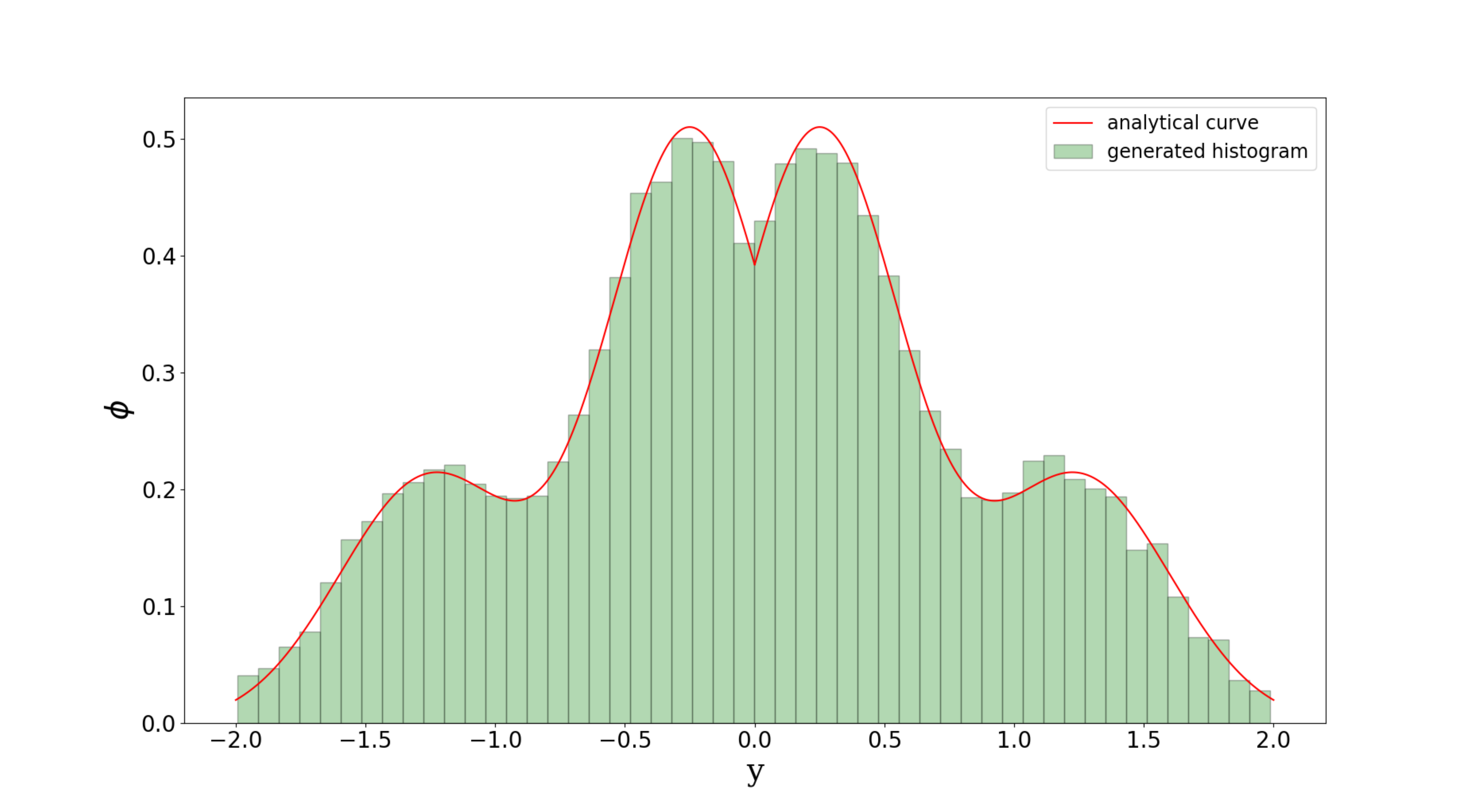}
    \caption{\footnotesize{Using the $1\,$M mesh}}
    \label{fig::ex_regen_b}
  \end{subfigure}
  \caption{Generation of an unweighted sample according to the 1D profile function at $\overline{x}=0$.
  The distribution is normalised to 1.}\label{fig::ex_regen}
\end{figure}

Since an example of the generation of the entire 2D sample has been already 
shown in section 2, here we focus on the constrained one-dimensional generation.   
Let us consider again the previous example and fix a $x$ value, for instance 
$\overline{x} = 0$. We generate $y$ values distributed as the profile function
\eqref{eq::profile} according to the above procedure.
In Figure~\ref{fig::ex_regen}, we plot the comparison between the generated points 
and the analytical profile function $\phi$ using our meshes with $100\,$k points 
(~\ref{fig::ex_regen_a}) and $1\,$M points  (~\ref{fig::ex_regen_b}). 
The generated histograms are in good agreement with the analytical curve and they 
reproduce well also the sharp deep in $y=0$. The result improves by exploiting 
the mesh with a greater number of points giving a solid indication that
the procedure is asymptotically converging to the true distribution.

\section{Systematic uncertainties}
\label{sysunc}
In summary, our fitting procedure is a way to approximate the probability density function associated to a given 2D scatter data with a piecewise function, i.e a histogram, with an automatic choice of the bins. In the example of the previous section, we have provided a non-trivial numerical proof of concept of it. Moreover, since in that case, the analytical distribution is known a-priori, we have a full control on the systematics uncertainties and the convergence of the method.

This is not the case in the practical applications, where the probability distribution is available only in the form of scatter data. As a consequence, estimating the systematics of the approximation becomes more difficult.  We follow a pragmatic approach which should provide a guideline for the user to tame the systematics according to his own scopes. Despite the fact that this systematics can be made smaller and smaller by providing more and more statistics (initial input events), in practice a compromise between the accuracy goal and the actual computational resources needs to be found. 

Let us start from some basic and general considerations. First, one can always separate the prediction of the total rates (including the geometrical acceptance of the detector) from those of more exclusive observables, as the angular distributions. Since the physical interaction cross section depends only on the energy, the inclusive total rates depend only on the effective dark sector particle energy distribution introduced in section~\ref{interaction} eq.\eqref{eq:flux_1D}. In our approach, we do not rely on the 2D fit to obtain this quantity. Instead, we perform a dedicated 1D fit exploiting a smoother class of functions. In this way we have a better control on the result and also on its uncertainties. Indeed, a 1D fit is a simpler operation and, since we are integrating over angles, we have access to a higher level of statistics.

The 1D fit works as follows. Starting from the input weighted data, we first build a 1D histogram and we assign to each energy bin the usual Poisson uncertainty. We then fit the histogram using a weighted cubic splines fitting. In order to assess the error on the fit, we add the possibility to vary the values within the histograms uncertainty bands. This can be done setting the parameter \texttt{rescale\_fac} in the \texttt{fit2D\_card} card file. which takes values in the interval $[-1,1]$, where 0 stands for the central value, 1 for the upper limit, -1 for the lower one. A realistic study case is given in the following subsection.  

We pass now to discussing the case of the 2D fit. At a fixed number of input events, the algorithm depends mainly on a unique parameters which gives the exit condition of the splitting loop. Naively, it represents the number of points which lie in each bin. Hence, there is a competition between choosing it small, to have a better description of the shape of the distribution (more bins with a smaller size), or choosing it large, to avoid to be overwhelmed by the statistical fluctuations (less bins with a larger size). The user can change this parameter by setting the value of the \texttt{npoints\_cell} variable in the \texttt{fit2D\_card} card file (the default value is 50). We have introduced in the code a consistency test (that can be enabled by setting to \texttt{True} the flag \texttt{fit\_syst}, again in the \texttt{fit2D\_card} card file) which compares the mesh obtained by varying the central value of this parameter by a factor of ${1}/{2}$ and a factor of $2$. To this aim, we consider the classifier 								
\begin{equation}
D(x) = \frac{P_1(x)}{P_1(x)+P_2(x)},
\end{equation}
where $x$ represents a generic event, and $P_i, i=1,2$, are the probability densities we are comparing. The values of $D$ ranges over the interval $[0,1]$. An average value $D \sim 1$ means that $P_2(x)$ underestimates $P_1(x)$, while for $D \sim 0$ we have the opposite. For $P_1(x)=P_2(x)$, $D(x)={1}/{2}$. Hence, in the case the average value $D \sim {1}/{2}$ and its standard deviation $\sigma_D$ is small, we cannot distinguish between the two probability functions.
The study of the $D$ classifier put on a quantitative foot the qualitative results given by the visual inspection of the mesh grids. Its mean and standard deviation give us a measure of the global goodness of the fit. Furthermore, we can perform also a more local comparison of the angular shape at fixed value of the energy variable. We postpone a more detailed discussion to the following subsection in which we apply the above analyses to a realistic case study.

\subsection{Study case: leptophobic model}
As a study case, we consider the example of the leptophobic model presented in the main part of this work. Since the generation of the input DM events can be simulated directly internally in MadDump, we have a direct access to input samples of different number of events. Our setup is outlined in the script reported in the listing appendix. We select the value $m = 2$ GeV for the mass of the DM mediator. We analyse first the uncertainties on the total rates. We reported the predictions for the DM yields in Tab.~\ref{tab::syst_total_rates_leptofobic_1} for input samples of increasing statistics. The uncertainties on the predictions correspond to the 1D fit variation around the histogram error bars, as stated above. The three results are consistent within their uncertainties and, as expected, the accuracy improves increasing the statistics. We observe, that in this case, for a sample of $100\,$k input events, which corresponds to $~100\,$k $ \times 0.28 = 28\,$k DM particles passing the geometrical acceptance, we already get a result accurate at the few percent level. Note that here we absorb the multiplicity factor of $2$ within the definition of the geometrical acceptance $\epsilon$
\begin{table}
\centering
\begin{tabular}{||c|c|c||}
	\hline
	\#evts & $\epsilon$ & \#DM\_evts \\ [0.5ex]
    \hline
    \hline
    $10\,$k   & $0.27$  & $25100^{+18\%}_{-15\%}$\\
    $100\,$k  & $0.282$ & $27700^{+4.9\%}_{-4.7\%}$ \\
    $1\,$M & $0.282$ & $28900^{+1.0\%}_{-1.7\%}$\\
    \hline
\end{tabular}
\caption{Estimates of the the total DM yields for increasing numbers of input events. The uncertainties refer to the variation around the 1D fit error bars as explained in the main text. The number of input here does not include the geometrical acceptance $\epsilon$, which is reported in the second column. }
\label{tab::syst_total_rates_leptofobic_1}
\end{table}	

The default value of the \texttt{npoints\_cell} parameter is $50$. We study what happens by varying it by a factor of $1/2$ and a factor of $2$. We briefly explain the strategy we followed for the comparison. We produce the three meshes corresponding to the three values $\texttt{npoints\_cell} = 25, 50, 100$. We use the result corresponding to the central value as our reference point and we denoted by $P(x)$ the corresponding probability density. This means that we evaluate $P$ starting from the 2D mesh as follows
\begin{equation}\label{meshpdf}
	P(x) = \frac{1}{n_\text{cells}}\frac{1}{A(x)}
\end{equation}
where $n_\text{cells}$ is the number of cells of the mesh and $A(x)$ is the area of the cells in which the point $x$ lies. We build the two classifiers
\begin{equation}
D_f(x) = \frac{P(x)}{P(x)+P_f(x)}, \qquad f={l,h}\,,
\end{equation}
where $P_l$ and $P_h$ are the probabilities densities corresponding respectively to the lower and the higher values of \texttt{npoints\_cell}. We generate random (uniformly distributed) points $ x = (E,\vartheta) $ and compute the mean and standard deviation of the two classifiers. We consider the simple unbiased estimator given by the uniform average
\begin{equation}
	\braket{D}_u = \frac{1}{N} \sum_{i=1}^N D(x_i), \qquad \sigma_u = \sqrt{\frac{1}{N-1} \sum_{i=1}^N [\braket{D}_u-D(x_i)]^2}.
\end{equation}
We reported our results in the third and fifth columns of Tab.~\ref{tab::syst_total_rates_leptofobic}, respectively for $D_l$ and $D_h$. All the mean values are fairly consistent with 0.5, which is the indication that the different meshes are consistent among themselves. The standard deviation is lower for $D_h$, which is what is indeed expected since, with a lower \texttt{npoints\_cell}, the fit is more sensitive to the statistical fluctuations of the original scatter data. Furthermore, we observe that the standard deviation decreases increasing the statistics from $10$\,k to $100$\,k events but, then, there are not any improvements from $100\,$k to $1\,$M. The explanation for this behavior is related to the vanishing tail of the distributions. Indeed, since we are fitting using piecewise functions, the accuracy of the method is worse in the long vanishing tails, where we decided to exploit a cut prescription instead of spreading uniformly the weights on a very big cell. Then, the error is dominated by the fluctuations near the boundary regions, where, however, the probability densities is approaching zero.
To test quantitatively this argument, we re-weight the events accordingly to our reference probability $P(x)$, and we consider the weighted estimators
\begin{equation}
	\braket{D}_w = \sum_{i=1}^N D(x_i) w(x_i), \qquad \sigma_w = \sqrt{\sum_{i=1}^N  [\braket{D}_w-D(x_i)]^2 w(x_i)},
\end{equation}
where the weights are given by
\begin{equation}
	w(x)=\frac{P(x)}{\int P(x) dx}.
\end{equation}
The results are reported in the fourth and sixth columns of Tab.~\ref{tab::syst_total_rates_leptofobic}. The errors drop significantly and for the largest input sample we see that it is not possible from the practical point of view to distinguish the probability densities given by the three meshes, so that one, for instance, might choose to use the mesh corresponding to \texttt{npoints\_cell}=25 since it is the finest (it has more cells wrt the other two).
\begin{table}
\centering
\begin{tabular}{||c|c|c|c|c|c||}
	\hline
	\#evts & $\epsilon$ & $ \braket{D_l}_u \pm \sigma_u $ & $\braket{D_l}_w \pm \sigma_w $ & $ \braket{D_h}_u \pm \sigma_u $ & $ \braket{D_h}_w \pm \sigma_w $ \\ [0.5ex]
    \hline
    \hline
    $10\,$k   & $0.27$  & $0.55 \pm 0.20$  & $0.54 \pm 0.03$ & $0.48 \pm 0.12$   & $0.549 \pm 0.013 $  \\
    $100\,$k  & $0.282$ & $0.54 \pm 0.12$  & $0.525 \pm 0.015$ & $0.50 \pm 0.08$   & $0.518 \pm 0.005 $   \\
    $1\,$M & $0.282$ & $0.55 \pm 0.13$  & $0.511 \pm 0.005$ & $0.47 \pm 0.09$   & $0.510 \pm 0.004 $   \\
    \hline
\end{tabular}
\caption{Comparison of the classifiers $D_f$, with $f=l,h$ calculated as weighted and unweighted averages.}
\label{tab::syst_total_rates_leptofobic}
\end{table}

Another important aspect concerns the convergence of our method to reproduce the original data set. With this, we mean the minimum number of regenerated events $N_\text{gen}$ needed to have a distribution which is consistent with that of the input data. Indeed, for $N_\text{gen}$ small, we expect the regenerated distribution is dominated by the statistical fluctuations. On the other hand, we expect that after having reached the desired level of agreement, further generations of events will not spoil the convergence. The naive expectation would be to have $N_\text{gen}^\text{min} \gtrsim N_\text{input}$. For a quantitative analysis, we rely again on the classifiers introduced above. In the following, we outline our strategy. We fix the mesh associated to the central value \texttt{npoints\_cell}=50 as our reference for the 2D-dimensional data distribution. Starting from this mesh, we regenerate samples of events with increasing statistics. We perform a second fit on top of each regenerated sample obtaining new meshes. We assume that these meshes represents the approximate bi-dimensional distributions of the samples, according to eq.~\eqref{meshpdf}. We finally compute the two averages for each of the classifier $D_f(x)$, where now $f=N_\text{gen}$ labels the regenerated samples. We report our result for the input sample of $100\,$k events. In this case, the effective number of input data events is $~30\,$k, i.e. the events passing the geometrical acceptance cuts of the detector. The results are reported in Tab.~\ref{tab::syst_ngen_convergence}. They confirm on the quantitative ground our naive expectations $N_\text{gen}^\text{min} \gtrsim N_\text{input}$, leading to the prescription $N_\text{gen}^\text{min} \sim 2-3 N_\text{input}$.
\begin{table}
\centering
\begin{tabular}{||c|c|c|||}
	\hline
	$N_\text{gen}$ & $\braket{D}_u \pm \sigma_u $ & $\braket{D}_w \pm \sigma_w $ \\ [0.5ex]
    \hline
    \hline
    $1\,$k  & $0.24 \pm 0.24$  & $0.27 \pm 0.23$ \\
    $3\,$k  & $0.31 \pm 0.20$  & $0.35 \pm 0.19$ \\
    $6\,$k  & $0.31 \pm 0.16$  & $0.34 \pm 0.17$ \\
    $10\,$k  & $0.38 \pm 0.16$  & $0.43 \pm 0.15$ \\
    $30\,$k  & $0.48 \pm 0.10$  & $0.50 \pm 0.11$ \\
	$60\,$k  & $0.48 \pm 0.10$  & $0.50 \pm 0.10$ \\
	$100\,$k  & $0.50 \pm 0.07$  & $0.51 \pm 0.08$ \\
	$300\,$k  & $0.54 \pm 0.07$  & $0.55 \pm 0.07$ \\
    \hline
\end{tabular}
\caption{Comparison of the classifiers $D$ (calculated as weighted and unweighted averages) as function of the regenerated number of events $N_\text{gen}$.}
\label{tab::syst_ngen_convergence}
\end{table}

\begin{figure}[h!]
  \centering
  \includegraphics[scale=0.7]{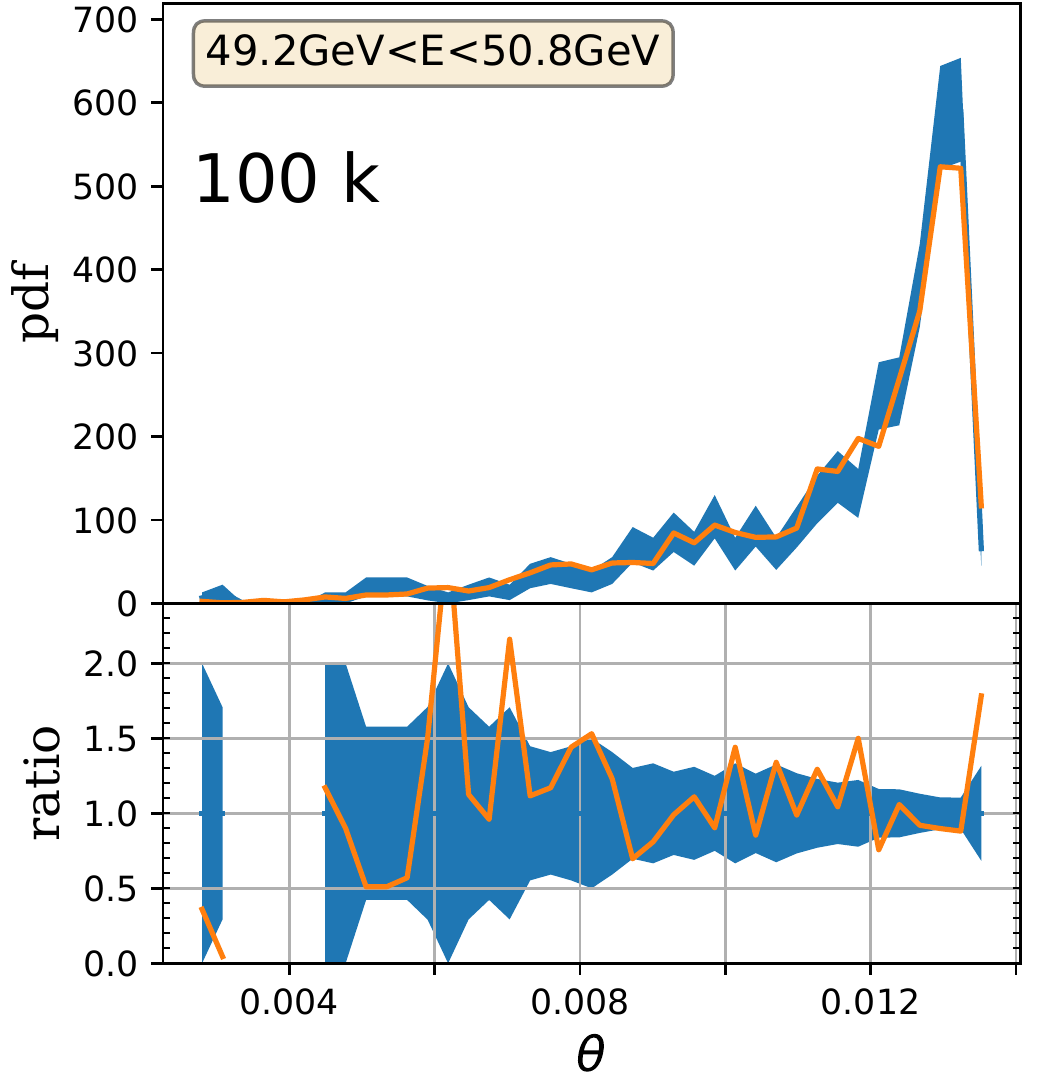}	
  \includegraphics[scale=0.7]{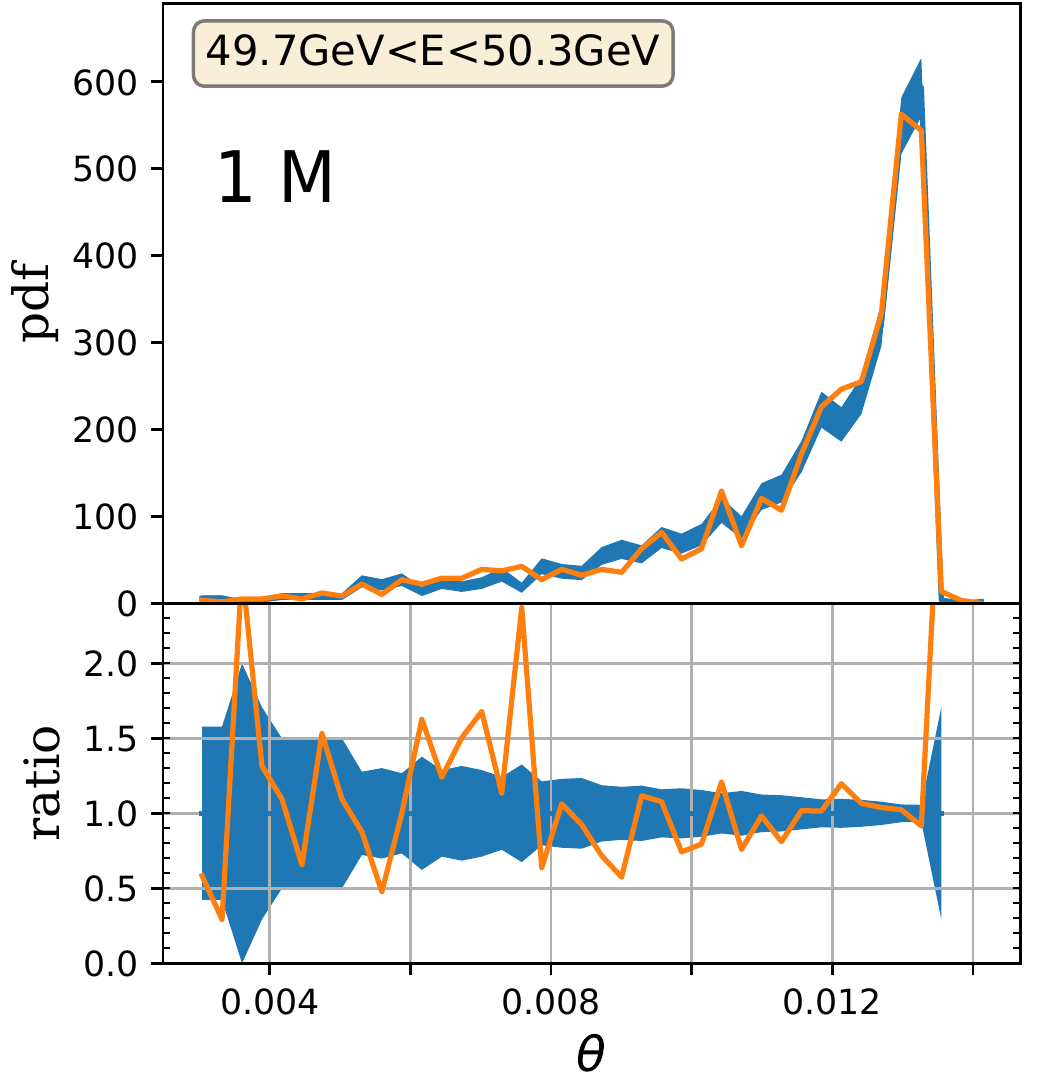}
  \includegraphics[scale=0.7]{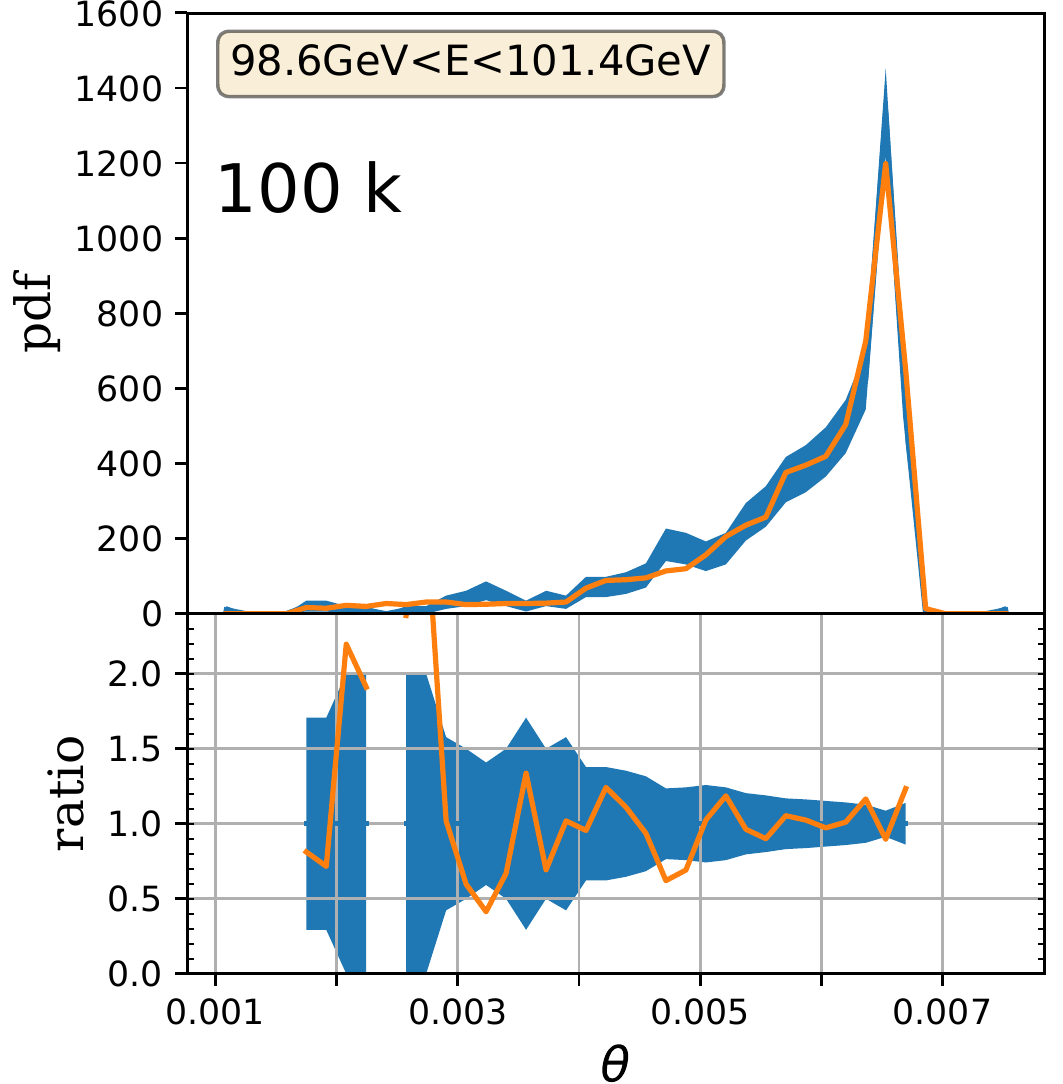}
  \includegraphics[scale=0.7]{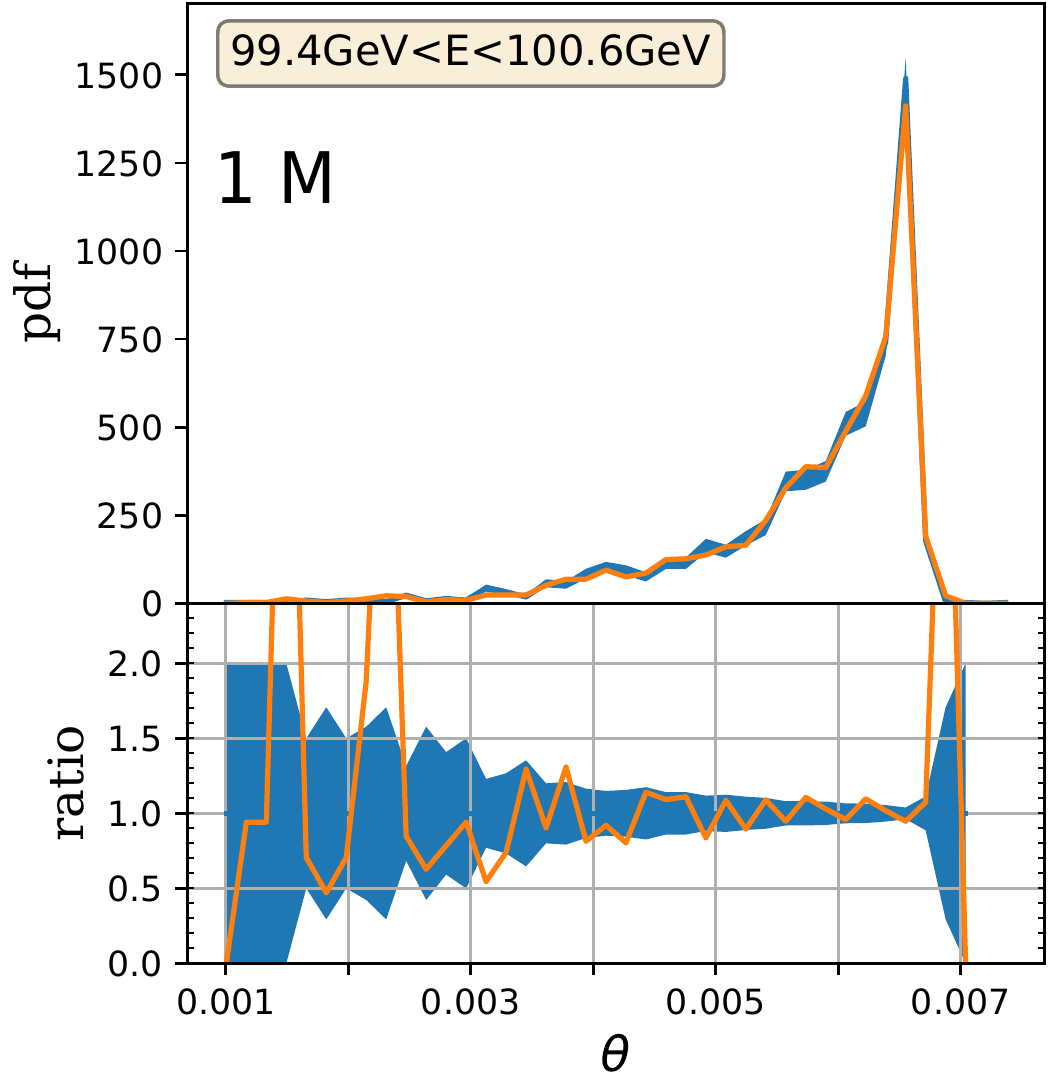}
  \caption{Normalized angular $\theta$-distributions for different energy bins for the case of $100\,$k inputs events (left panels) and of $1\,$M inputs events (right panels). The blue band represents the original data with the associated Poissonian uncertainties. The orange line represents the histogram of the points regenerated with the 2D mesh. The energy bins size is automatically determined by MadDump in a such a way to have comparable amount of statistics in each bin. }
  \label{fig:angular_distributions-100k}
\end{figure}

We conclude this discussion on the fit systematics showing some lesser inclusive results at the level of differential distributions wrt to the polar angle.
We adopted the following strategy. We select an energy value $E$. Then, we pick the cells in the 2D mesh fit in which $E$ is included.  
We consider as energy resolution parameter some multiple (the default is 2) of the minimum energy width of the selected cells. This allows us to consider an energy bin centered in $E$ with width given by the resolution parameter.
Then, we consider the input data and the regenerated ones which lie in this bin and we compared their angular distributions given by standard 1D-histogram. Our choice of the energy bin size guarantees that the statistics is comparable for any starting $E$ values.
We have analyzed and compared the results obtained with the two input samples $100\,$k and $1\,$M. They are shown in Fig.~\ref{fig:angular_distributions-100k} and Fig.~\ref{fig:angular_distributions-1M}, respectively. In both cases, we have used regenerated samples with $N_\text{gen}=300\,$k. We found a good agreement both between data distributions for different statistics and between data and our fits which is of the order of $\sim 25-30\%$ for the $100\,$k case and $\sim 10\%$ for the $1\,$M one. As expected, in the region corresponding to the bulk of the events, the corresponding energy bin size are smaller. For example, in our study case, we observe that in the central energy range $ 40\,$ GeV $< E < 150\,$GeV the energy bin size is of order $\sim 1\,$GeV for the $100\,$k input sample and the situation fairly improves for the $1\,$M one. On the contrary, for the value $E=20\,$GeV which lies on the tail of the distribution, we need a bigger bin size, $\sim 16-18\,$GeV.

The analyses performed in this section are encoded in MadDump and the user can reproduce the same studies for his particular situation. As a rule of thumb, to take \textit{cum grano salis}, $500\,$k events entering the detector can be consider a reasonable amount of input statistics. Together with the default settings of the internal MadDump parameters and the choice of $N_\text{gen}^\text{min} \sim 2-3 N_\text{input}$ this should lead to an uncertainty of $1\%$ for the total rates and $5-10\%$ on the angular distributions, which is usually lesser than the other systematics of the simulation. 

\section{A method to take into account the depth of the detector}
\label{Details2}
Consider the scattering of a flux of incident particles impinging on a 
thick target. Let us fix the geometry of the problem and consider for instance 
a parallelepiped shape for the fiducial volume of the target.
In general, the incident flux is neither collimated nor mono-energetic. For the sake of simplicity, we consider the flux to be originating from a point-like source placed on the target axis and mono-energetic. 
Consider a cartesian 3D reference frame in which the z-axis is along the ``depth'' of
the target and x and y are parallel to the other dimensions. Let us subdivide the target in thin sheets along the z-axis. The number of
scattered events in each sheet is given by 
\begin{equation}
 dN(z) = \int_{S(z)} dx dy \frac{\partial^2 F}{\partial x \partial y}(x,y,z) \rho(x,y,z) 
 \sigma(z) dz \,,
\end{equation}
where $S(z)$ is the surface of the shell at $z$, F is the flux of incident 
particles impinging on the sheet, $\rho$ is the number density of the target 
particles and $\sigma$ is the interaction cross section. In the formula above,
we assumed that the cross section is constant all over the surface 
of the sheet. 
Furthermore, we assume that the cross section is constant over the whole
fiducial volume of the target and we consider a uniform target.
Then, integrating also over the depth of the target, we get
the master formula for the number of scattered events
\begin{equation}
 N = \rho \sigma \int_{z_0}^{z_1} dz \int_{S(z)} dxdy \frac{\partial^2 F}{\partial x \partial y}(x,y,z).
 \end{equation}  
We are interested in the situation in which the cross section is very small,
i.e. we can neglect (at least in first approximation) 
the variation of the flux $F$ due to the scattered particles. This means
that the dependence of ${\partial^2F }/{\partial x \partial y}$ on $z$ is
purely geometrical: from a given  configuration at a point $z^*$ 
it is possible to construct the flux at a new $z$ point by prolongating the
flying direction of the particles in the flux. For this reason, in the 
above example, even though the surface of the sheet is constant, the number of 
particles impinging on the different sheets along the z-axis is different:
\begin{equation}\label{ndef}
  \int_{S} dxdy \frac{\partial^2 F}{\partial x \partial y}(x,y,z) = \overline{n}(z)\,.
\end{equation}
In Monte Carlo integration/generation this translates in employing different
weights for the bunch of events describing the incident flux. It is still
possible to use unweighted events if one considers a variant of the hit-or-miss
rejection method. Here, by ``unweighted events'' we mean that 
the points have been generated according to the 
$dxdy \frac{\partial^2 F}{\partial x \partial y}$ distribution. 
Since the integrand is positive definite, enlarging the integration region 
we obtain the inequality
\begin{equation}
 S(z) \ge S \implies   \int_{S} dxdy \frac{\partial^2 F}{\partial x \partial y}(x,y,z) < \int_{S(z)} dxdy \frac{\partial^2 F}{\partial x \partial y}(x,y,z) \equiv n^*(z)\,.
\end{equation}
In particular, we can choose $S(z)$ such that the above integral $n^*$ is constant.
This means that we enlarge the surface according to the radial projection
starting from the point-source of the flux. Then, for the new integral, we can
employ unweighted events for the flux: 
\begin{equation}
 N = \rho \sigma n^* \int_{z_0}^{z_1} dz = \rho \sigma n^*(z_1-z_0)\,.
\end{equation}
A full event is given once a $z$ variable or equivalently a travel distance 
along the flying direction of the event is generated uniformly between the 
minimum and the maximum value inside the largest volume. 
Then, we accept or reject the event whether it lies or not in the true  fiducial volume. 

\par While correct, this method is not efficient as generated events can be rejected. An alternative approach entails applying a simple reweighting procedure.
Intuitively, we just have to penalize the events that would be produced by particles crossing the fiducial volume of the detector over smaller paths.
Indeed, a given event may contribute
or not to $n(z)$ in Eq.~\eqref{ndef} depending whether at $z$ 
it is inside or not the integration region:
\begin{equation}
\frac{\partial^2 F}{\partial x \partial y}(x,y,z) = 
\begin{cases}
  \frac{\partial^2 F}{\partial x \partial y}(x(z),y(z),z_0), z<d(x,y)\\
  0, z>d(x,y)\\
\end{cases}  
\end{equation}
where we denoted with $d(x,y)$ the z-distance after which the event 
goes out of the integration region. However, the weights retain a dependence on 
$z$ due to the presence in the argument of the function of ($x(z)$,$y(z)$),
which represent the coordinates of the particle when it crossed the sheet 
at $z_0$. If we replace $x,y$-coordinates by angular ones $(\theta,\phi)$ 
which gives the flying direction of the events,
the weight will not have any residual dependence on $z$ but the theta function 
$\Theta(z-d(\theta,\phi))$. 
This can be simply taken into account by 
reweighting the events as 
\begin{equation}
  \frac{\partial^2 F}{\partial x \partial y}(\theta,\phi) \times \frac{d(\theta,\phi)}{(z_1-z_0)}\,.
\end{equation} 
In terms of the travel distance inside the fiducial volume of the detector $r(\theta,\phi)$, we have
\begin{equation}
 d(\theta,\phi) = r(\theta,\phi)\cos(\theta).
\end{equation}
Then, using this reweighting strategy, we can reconstruct the full event by generating uniformly the $z$ or, equivalently, the traveled distance variable 
according to the actual minimum-maximum allowed by the geometry of the target.

Notice that the number of scattered events in the two cases is given by:
\begin{equation}
 N = \begin{cases}
   \rho \sigma (z_1-z_0) (n^*-n_\text{rejected}) \\
   \rho \sigma (z_1-z_0) \langle\overline{n}(z)\rangle,
\end{cases}
\end{equation}
which implies the integral condition:
\begin{equation}
   (n^*-n_\text{rejected}) = \langle\overline{n}(z)\rangle.
\end{equation}

\section{Listings}
\label{Listings}
\lstset{basicstyle=\scriptsize\ttfamily,
  keywordstyle=\color{Orchid}\bfseries,
  commentstyle=\color{Red},
  stringstyle=\color{BrickRed},
  showstringspaces=true,
  columns=fullflexible,
  frame = single,
}

In this appendix we report the input script files used for the examples presented in the main text.  

\subsection{Leptophobic GeV mediator}
\begin{lstlisting}
import model DMZB
generate production p p > chidmsc chidmsc~ 
define darkmatter chidmsc
add process interaction @DIS 
output leptofobic
launch
set nevents 100k
set ebeam1 400.
set ebeam2 0.938
set use_syst False
set flux_norm 19663072216.4
set prod_xsec_in_norm True
set d_target_detector 5650.0
set detector_density  3.72
set parallelepiped True
set x_side 187.0
set y_side 69.0
set depth 200.0
set testplot True
set mchidm 75
set mchidmsc 0.75
set mzb scan:[i for i in range(2,11,1)]
set wzb auto
\end{lstlisting}

\subsection{Scalar GeV mediator}
\begin{lstlisting}
import model DMsimp_UFO-full
generate production p p > xd xd~ /y1
define darkmatter xd
add process interaction @DIS /y1
output scalar
launch
set nevents 100k
set ebeam1 400.
set ebeam2 0.938
set use_syst False
set flux_norm 19663072216.4
set prod_xsec_in_norm True
set d_target_detector 5650.0
set detector_density  3.72
set parallelepiped True
set x_side 187.0
set y_side 69.0
set depth 200.0
set testplot True
set gsxr 0.0
set gsxc 0.0
set gsxd 1.0
set gpxd 0.0
set gsd11 1e-3
set gsu11 1e-3
set gsd22 0.0
set gsu22 0.0
set gsd33 0.0
set gsu33 0.0
set gpd11 0.0
set gpu11 0.0
set gpd22 0.0
set gpu22 0.0
set gpd33 0.0
set gpu33 0.0
set gsg 0.0
set gpg 0.0
set gvxc 0.0
set gvxd 0.0
set gaxd 0.0
set gpxd 0.0
set gvd11 0.0
set gvu11 0.0
set gvd22 0.0
set gvu22 0.0
set gvd33 0.0
set gvu33 0.0
set gad11 0.0
set gau11 0.0
set gad22 0.0
set gau22 0.0
set gad33 0.0
set gau33 0.0
set mxd 0.75
set my0 scan:[i for i in range(2,11,1)]
set wy1 auto
set ymdo 2.462206e+02 
set ymup 2.462206e+02 
\end{lstlisting}

\subsection{DP from pion decays}
\begin{lstlisting}
import model DM_mesons_2
#import the input file events "MesonFulx.hepmc"
import_events decay ./MesonFlux.hepmc
decay pi0 > y1 a, y1 > xd xd~
define darkmatter xd
add process interaction @DIS
add process interaction @electron
output DP_electron
launch
set flux_norm 2.0e20
set prod_xsec_in_norm false 
set d_target_detector 5650.0
set detector_density 3.72
set Z_average 82
set A_average 207
set parallelepiped True
set x_side 187.0
set y_side 69.0
set depth 200.0
set ncores 16
set testplot True
set gvd11 -3.333333e-4
set gvu11  6.666666e-4
set gvd22 -3.333333e-4
set gvu22  6.666666e-4
set gvd33 -3.333333e-4
set gvu33  6.666666e-4
set gvl11 -1.000000e-3
set gvl22 -1.000000e-3
set gvl33 -1.000000e-3
set my1 scan1:[0.01*i for i in range(1,14)]
set mxd scan1:[0.01/3.*i for i in range(1,14)]
set wy1 auto
\end{lstlisting}
  
\bibliographystyle{JHEP}
\bibliography{biblio}

\end{document}